\documentclass[apj]{emulateapj}
\usepackage{graphicx}

\newcommand{\hi}          {\mbox{\rm \ion{H}{1}}}
\newcommand{\hii}         {\mbox{\rm \ion{H}{2}}}
\newcommand{\htwo}        {\mbox{H$_{2}$}}
\newcommand{\jone}        {$J=1\rightarrow0$}

\newcommand{\ha}          {H$\alpha$}
\newcommand{\kmpers}      {\mbox{\rm km~s$^{-1}$}}

\newcommand{\msun}        {\mbox{\rm M$_\odot$}}
\newcommand{\msunperpcsq} {\mbox{\rm M$_\odot$~pc$^{-2}$}}
\newcommand{\msunperpcsqyr} {\mbox{\rm M$_\odot$~pc$^{-2}$~yr$^{-1}$}}
\newcommand{\msunperyr}   {\mbox{\rm M$_\odot$~yr$^{-1}$}}
\newcommand{\msunperpccu} {\mbox{\rm M$_\odot$~pc$^{-3}$}}
\newcommand{\msunperyrkpcsq} {\mbox{\rm M$_\odot$~yr$^{-1}$~kpc$^{-2}$}}
\newcommand{\xco}         {\mbox{$X_{\rm CO}$}}
\newcommand{\aco}         {\mbox{$\alpha_{\rm CO}$}}
\newcommand{\xcounits}    {\mbox{\rm cm$^{-2}$(K km s$^{-1}$)$^{-1}$}}
\newcommand{\acounits}  {\mbox{\rm M$_\odot$ (K km s$^{-1}$ pc$^2$)$^{-1}$}}
\newcommand{\Lcounits}  {\mbox{\rm K km s$^{-1}$ pc$^2$}}
\newcommand{\Kkmpers}     {\mbox{\rm K km s$^{-1}$}}
\newcommand{\Kkmperspcsq} {\mbox{\rm K km s$^{-1}$ pc$^2$}}

\newcommand{\Ico}         {\mbox{I$_{\rm CO}$}}

\newcommand{\percmsq}     {\mbox{cm$^{-2}$}}
\newcommand{\cii}         {\mbox{\rm [\ion{C}{2}]}}

\newcommand{\Smol}	  {\mbox{$\Sigma_{\rm mol}$}}
\newcommand{\Mmol}	  {\mbox{${\rm M}_{\rm mol}$}}
\newcommand{\Ssfr}	  {\mbox{$\Sigma_{\rm SFR}$}}
\newcommand{\Sgas}	  {\mbox{$\Sigma_{\rm gas}$}}
\newcommand{\Sgbc}	  {\mbox{$\Sigma_{\rm gbc}$}}
\newcommand{\Sdiff}	  {\mbox{$\Sigma_{\rm diff}$}}
\newcommand{\Shi}	  {\mbox{$\Sigma_{\rm HI}$}}
\newcommand{\Shtwo}	  {\mbox{$\Sigma_{\rm H2}$}}
\newcommand{\Sha}	  {\mbox{$\Sigma_{\rm H\alpha}$}}
\newcommand{\SFE}	  {\mbox{\rm SFE}}
\newcommand{\fw}	  {\mbox{$f_{\rm w}$}}
\newcommand{\dgdr}         {\mbox{$\delta_{\rm GDR}$}}
\newcommand{\Pth}         {\mbox{$y$}}

\begin{document}

\title{The State of the Gas and the Relation Between Gas and Star Formation at Low Metallicity: the Small Magellanic Cloud}

\author{Alberto D. Bolatto\altaffilmark{1}, Adam K. Leroy\altaffilmark{2,13},
  Katherine Jameson\altaffilmark{1}, Eve Ostriker\altaffilmark{1},
  Karl Gordon\altaffilmark{3}, Brandon Lawton\altaffilmark{3}, Sne\v{z}ana
  Stanimirovi\'c\altaffilmark{4}, Frank P. Israel\altaffilmark{5},
  Suzanne C.  Madden\altaffilmark{6}, Sacha Hony\altaffilmark{6},
  Karin M. Sandstrom\altaffilmark{7,14}, Caroline Bot\altaffilmark{8},
  M\'onica Rubio\altaffilmark{9}, P. Frank Winkler\altaffilmark{10},
  Julia Roman-Duval\altaffilmark{3}, Jacco Th. van
  Loon\altaffilmark{11}, Joana M. Oliveira\altaffilmark{11}, and
  R\'emy Indebetouw\altaffilmark{2,12}}

\altaffiltext{1}{Department of Astronomy and Laboratory for
  Millimeter-wave Astronomy, University of Maryland, College Park, MD
  20742, USA}
\email{bolatto@astro.umd.edu}

\altaffiltext{2}{National Radio Astronomy Observatory, Charlottesville, VA 22903, USA}

\altaffiltext{3}{Space Telescope Science Institute, Baltimore, MD 21218, USA}

\altaffiltext{4}{Department of Astronomy, University of Wisconsin, Madison, WI 53706, USA}

\altaffiltext{5}{Leiden Observatory, Leiden, NL-2300 RA, The Netherlands}

\altaffiltext{6}{Laboratoire AIM Paris-Saclay, CNRS/INSU CEA/Irfu Universit\'e Paris Diderot, 91191 Gif sur Yvette, France}

\altaffiltext{7}{Max-Planck-Institut f\"ur Astronomie, D-69117 Heidelberg, Germany}

\altaffiltext{8}{Universit\'e de Strasbourg Observatoire Astronomique de Strasbourg and CNRS, 67000 Strasbourg, France}

\altaffiltext{9}{Departamento de Astronom\'{\i}a, Universidad de Chile, Casilla 36-D, Santiago, Chile}

\altaffiltext{10}{Department of Physics, Middlebury College, Middlebury,
VT 05753, USA}

\altaffiltext{11}{Lennard-Jones Laboratories, Keele University, ST5 5BG, UK}

\altaffiltext{12}{Department of Astronomy, University of Virginia, Charlottesville, VA 22904, USA}

\altaffiltext{13}{Hubble Fellow}

\altaffiltext{14}{Marie Curie Fellow}

\begin{abstract}
We compare atomic gas, molecular gas, and the recent star formation
rate (SFR) inferred from \ha\ in the Small Magellanic Cloud (SMC). By
using infrared dust emission and local dust-to-gas ratios, we
construct a map of molecular gas that is independent of CO
emission. This allows us to disentangle conversion factor effects from
the impact of metallicity on the formation and star formation
efficiency of molecular gas. On scales of 200~pc to 1~kpc (where the
distributions of \htwo\ and star formation match well) we find a
characteristic molecular gas depletion time of $\tau_{\rm dep}^{\rm
mol} \sim 1.6$~Gyr, similar to that observed in the molecule-rich
parts of large spiral galaxies on similar spatial scales. This
depletion time shortens on much larger scales to $\sim 0.6$~Gyr
because of the presence of a diffuse \ha\ component, and lengthens on
much smaller scales to $\sim 7.5$~Gyr because the \ha\ and \htwo\
distributions differ in detail. We estimate the systematic
uncertainties in our dust-based $\tau_{\rm dep}^{\rm mol}$ measurement
to be a factor of $\sim 2$--$3$. We suggest that the impact of
metallicity on the physics of star formation in molecular gas has at
most this magnitude, rather than the factor of $\sim 40$ suggested by
the ratio of SFR to CO emission. The relation between SFR and neutral
($\htwo+\hi$) gas surface density is steep, with a power-law index
$\approx2.2\pm0.1$, similar to that observed in the outer disks of
large spiral galaxies.  At a fixed total gas surface density the SMC
has a $5-10$ times lower molecular gas fraction (and star formation
rate) than large spiral galaxies. We explore the ability of the recent
models by
\citet{KRUMHOLZ09} and \citet{OSTRIKER10} to reproduce our
observations. We find that to explain our data at all spatial scales
requires a low fraction of cold, gravitationally-bound gas in the
SMC. We explore a combined model that incorporates both large scale
thermal and dynamical equilibrium and cloud-scale photodissociation
region structure and find that it reproduces our data well, as well as
predicting a fraction of cold atomic gas very similar to that observed
in the SMC.
\end{abstract}


\section{Introduction}

The relation between gas content and star formation activity in
galaxies has been a matter of intense investigation since the
pioneering work of Maarten Schmidt. \citet{SCHMIDT59} suggested that
the star formation rate (SFR) in a galaxy is proportional to a power
of the gas density, such that $\rho_{\rm SFR}\sim\rho_{\rm
gas}^n$, where $n\simeq1-3$ and most likely $n=2$ based on a number of
arguments that included the luminosities of open clusters, the
abundance of helium, and the vertical distribution of objects in the
plane of the Milky Way.  Here we refer to the quantitative relation
between gas and star formation as the ``star formation law'' for
convenience without intending to suggest a physical law or a specific
functional form.

Most modern empirical studies of the extragalactic star formation law
follow those by \citet{KENNICUTT89,KENNICUTT98}, who studied
disk-averaged correlations in a sample of $\sim100$ galaxies including
starbursts and high-mass dwarfs. This work linked the surface density
of star formation rate, $\Ssfr$, to the surface density of total neutral
(atomic and molecular) gas, $\Sgas=\Shi+\Shtwo$. \citet{KENNICUTT98}
found that $\Ssfr\propto\Sgas^{1+p}$ with $1+p\approx {1.4\pm0.15}$
for his composite sample of galaxies. Since the gas depletion time
$\tau_{\rm dep}^{\rm gas}\equiv \Sgas/\Ssfr \propto \Sgas^{-p}$ for a
power-law relation, $p=0$ indicates that the gas depletion time is
constant (independent of environment), while $p>0$ indicates that star
formation is more rapid in high-$\Sgas$ regions. In practice, it is
also important to keep in mind the observational complexities
associated with measuring \Sgas\ and particularly \Shtwo\ from CO
observations, and the systematics thus introduced.

Several subsequent studies focused on the star formation law
within galaxies, employing high resolution molecular and
\hi\ observations \citep{MARTIN01,WONG02,BOISSIER03}. More recently,
\citet{BIGIEL08} and \citet{LEROY08} analyzed 12 nearby spirals
with high quality \hi, CO, far-infrared, and far-ultraviolet data. These
observations show a clear correlation between $\Ssfr$ and the surface
density of molecular gas, $\Shtwo$, with an approximately constant
star formation rate per unit molecular gas, yielding an approximately
linear power-law index $1+p=1.0\pm0.2$. By contrast, they found a
steep correlation between $\Ssfr$ and the surface density of atomic
gas, $\Shi$ \citep[the observed correlation in the optical disks of
large galaxies has $1+p\gtrsim 2.7-3.5$, and $1+p\sim2$ in the
\hi-dominated outer disks][]{BIGIEL08,BIGIEL10b}. \citet{SCHRUBA11} and
\citet{BIGIEL11} extended these findings to a larger sample of 30 disk
galaxies, while \citet{BLANC09} and \citet{RAHMAN11} arrived at similar
conclusions in very detailed studies of individual targets.

A consequence of the observed linearity of the star formation law over
the molecule-dominated regions of disks is that the typical time scale
to deplete the molecular gas by star formation in a disk galaxy is
$\tau_{\rm dep}^{\rm mol}=\Smol/\Ssfr\sim1.9\pm0.9$ Gyr (\Smol\
corresponds to \Shtwo\ corrected by a 1.36 factor to account for the
cosmic abundance of helium).  The lack of dependence of $\tau_{\rm
dep}^{\rm mol}$ on environment can be naturally understood if two
conditions are fulfilled.  The first condition is for star formation
to be a local process primarily determined by the conditions inside
the giant molecular clouds (GMCs) where it takes place. GMCs are
themselves isolated from the global galactic environment provided that
they are self-gravitating and therefore overpressured with respect to
the neighboring gas \citep{MO07}. The second condition is for the
properties of GMCs to be universal, and therefore independent of their
galactic environment. Because star formation takes place in the
densest regions of GMCs, themselves self-gravitating and thus mostly
decoupled from their surroundings, the first condition appears likely
\citep{KRUMHOLZ05}, at least in mid-disk and outer-galaxy regions.  In
galactic centers it is less clear whether there are isolated,
gravitationally bound GMCs, or instead a distributed molecular medium.
Environmental considerations may be more important if cloud--cloud
collisions play an important role \citep{TAN00}. Similarly, the second
condition appears to be broadly satisfied as the resolved properties
of GMCs (or at least of the dense regions of GMCs, bright in CO) are
observed to be remarkably constant over a wide range of galaxy
environments (the Local Group spirals and nearby dwarfs, Bolatto et
al. 2008\nocite{BOLATTO08}; the Milky Way, Heyer et
al. 2009\nocite{HEYER09}; the LMC, Hughes et al. 2010\nocite{HUGHES10};
the outer disk of M33, Bigiel et al. 2010\nocite{BIGIEL10a}).

Most of the studies of the relation between gas and star formation,
and the ensuing conclusions, focus on large galaxies which tend to be
rich in molecules. There is a dearth of comparable information for
low-mass low-metallicity star-forming dwarf galaxies, primarily
because such objects emit only very faintly in molecular gas tracers
such as CO. The measurements that do exist reveal large ratios of SFR
to CO emission in late-type, low-mass galaxies
\citep[e.g.,][]{YOUNG96,GARDAN07,LEROY07B,KRUMHOLZ11}.

Studies of the star formation law in dwarf galaxies are interesting
for a number of reasons. Fundamentally they probe a very different
physical regime from large spiral galaxies, one where atomic gas
dominates the surface density of the interstellar medium (ISM) on
large scales, and elements heavier than helium are less abundant. This
dearth of heavy elements causes a number of differences in gas
chemistry and physical conditions. For example, dust-to-gas ratios are
lower, producing lower extinctions and higher photodissociation rates
\citep[the likely cause of their general CO
faintness,][]{ISRAEL86,MALONEY88,LEQUEUX94,BOLATTO99}.

More importantly, studies of the star formation law in low-mass
low-metallicity star-forming dwarf galaxies provide fertile testing
ground for star formation theories.  A natural implication of the
approximately constant molecular $\tau_{\rm dep}^{\rm mol}$ among
normal galaxies is that the effectiveness at forming molecular gas
plays a critical role in establishing the star formation rate. The
molecular gas fraction varies systematically both within and among
galaxies \citep{YOUNG91,WONG02,BLITZ06,LEROY08}. 
Dwarf galaxies appear to have low molecular fractions, despite their
large gas fractions and long time scales to deplete gas reservoirs at
current rates of star formation. If their $\tau_{\rm dep}^{\rm mol}$
values are similar to normal galaxies, it would suggest that GMC formation
(and destruction) is the rate-limiting step for star formation in
dwarf systems.

Because their low metallicities contrast with those in spiral
galaxies, star-forming dwarf galaxies can break a number of
degeneracies in the physical drivers of the molecular
fraction. Different models predict very different behaviors for the
impact of metallicity on molecular fractions and star formation
rates according to their emphasis on dynamics, thermodynamic
equilibrium, or dust shielding.  For example, the requirement of a
minimum extinction for star formation to occur
\citep[e.g.,][]{MCKEE89,LADA09} would have a proportionally more
important impact in low metallicity objects, while models that are
solely based on dynamics cannot distinguish between low and high
metallicity.
 
The main obstacle to using dwarf galaxies to test various aspects of
the star formation law is the difficulty in observing their molecular
gas distribution, stemming from the faintness of their CO emission and
the uncertainty in its quantitative relation to \htwo. The Small
Magellanic Cloud (SMC) offers a unique opportunity in this regard. Its
proximity allows us to probe small spatial scales with observations of
modest angular resolution. As a result, we can use dust emission to
construct a map of $\Sigma_{\rm H2}$ that does not depend on CO. In
this study we use such a map to compare the distributions of molecular
gas, atomic gas, and recent star formation traced by ionizing photon
production.

\subsection{The Small Magellanic Cloud as a Galaxy}
\label{subsec:smcintro}

By virtue of its location and properties the SMC is a prime laboratory
for the study of the relation between gas and star formation in
low-mass galaxies. Indeed, the SMC was the target of the first
extragalactic study of the star formation law, relating \hi\ and
stellar surface densities \citep{SANDULEAK69}.  Located scarcely
61~kpc away \citep{HILDITCH05,KELLER06,SZEWCZYK09}, the SMC is the
nearest gas-rich low metallicity dwarf galaxy with active
star-formation \citep[$Z_{\rm SMC}\sim
Z_\odot/5$,][]{DUFOUR84,KURT99,PAGEL03}. As such, it provides invaluable
insight into the physics and chemistry of the ISM in chemically
primitive star-forming galaxies. Moreover, because of its proximity it
is possible to carry out studies on the scale of individual young
stellar objects, which reveal subtle differences in the temperature
and chemistry of star-forming cores \citep{VANLOON10b,OLIVEIRA11}.

With an \hi\ mass of $M_{\rm HI}\simeq4.2\times10^8$~\msun\
\citep{STANIMIROVIC99} the SMC is rich in atomic gas,
although also strikingly defficient in cold \hi\
\citep{DICKEY00}. It was already apparent in the early observations
that the distribution of the \hi\ is complex, with more than one
component along the line-of-sight \citep{MCGEE81}. This complexity is
at least in part due to the presence of several supergiant shells
\citep{STANIMIROVIC99}. The underlying dynamics appear
to be disk-like with an inclination $i\approx40^\circ\pm20^\circ$,
although disturbed by the interaction with the Milky Way and the LMC
\citep{STANIMIROVIC04}. In our analysis we will ignore this
complexity, but the geometry of the SMC and the level of turbulent
support in the gas remain some of the most important caveats in the
comparison to star formation models.

The SMC hosts a healthy amount of star formation despite being
disproportionately faint in CO emission
\citep{KENNICUTT95,ISRAEL86,ISRAEL93}.  In fact, in the SMC the CO is
under-luminous with respect to the star formation activity by almost
two orders of magnitude when compared to normal disks, or even more
massive small galaxies.  A possible explanation for this fact is that
the SMC is extraordinarily efficient at turning the available
molecular gas into stars (i.e., the H$_2$ depletion time is very
short), which would suggest that we are either observing an
out-of-equilibrium situation (a fleeting starburst) or that the
conditions conspire to keep a small reservoir of extremely short-lived
GMCs.  A more likely alternative is that the weak CO emission is not
representative of star-forming molecular gas, with H$_2$/CO
considerably larger than in the Milky Way
\citep{ISRAEL86,RUBIO93b}.

\citet{LEROY07} used new far-infrared observations
to map the H$_2$ distribution in the SMC bypassing CO emission,
following an extension of the methodology previously employed by
\citet{ISRAEL97} in the Magellanic Clouds. The feasibility of using
dust to map \htwo\ was clearly demonstrated by \citet{DAME01} in the
Milky Way \citep[see also][for an early study of the correlation
between dust emission and gas in the Milky Way]{BLOEMEN90}. It was
also shown to be consistent with virial masses on large scales
\citep{LEROY09}. Furthermore, a systematic application of an extension
of this methodology throughout the Local Group produces molecular
masses that are consistent with those obtained by other methods at
approximately Galactic metallicities \citep{LEROY11}.  The analysis by
\citet{LEROY07} shows that the column density of molecular gas present
in the SMC is far in excess of that derived by applying the Galactic
CO-to-H$_2$ conversion factor, \xco, to the CO observations.  Their
modeling yields a total molecular mass $M_{\rm
H2}\approx3.2\times10^7$ M$_\odot$. Therefore the neutral ISM in the
SMC is approximately 10\% molecular, a lower fraction than observed in
normal late spirals but not dramatically so
\citep{YOUNG91}. Further refinements to the analysis \citep[][and this
work]{LEROY09,LEROY11} broadly confirm these numbers.

In this study we analyze the spatially resolved correlation between
the distributions of molecular gas, atomic gas, and recent star
formation traced by ionizing photon production on different scales,
comparing the star formation law in this small low-metallicity galaxy
with that in large disks. We present our data and discuss the
methodology we use to measure molecular column densities and star
formation rates in \S\ref{sec:method}. We show our main observational
results in \S\ref{sec:results}, focusing on the relation between
\htwo\ and star formation in \S\ref{subsec:molsfr}, total gas and star
formation in \S\ref{subsec:gassfr}. In \S\ref{sec:comparison}, we
compare our results to recent analytical physical models of star
formation in galaxies. Finally, we summarize our conclusions in
\S\ref{sec:conclusions}.

\section{Observations and Methodology}
\label{sec:method}

\begin{figure*}[t!]
\epsscale{0.8}
\plotone{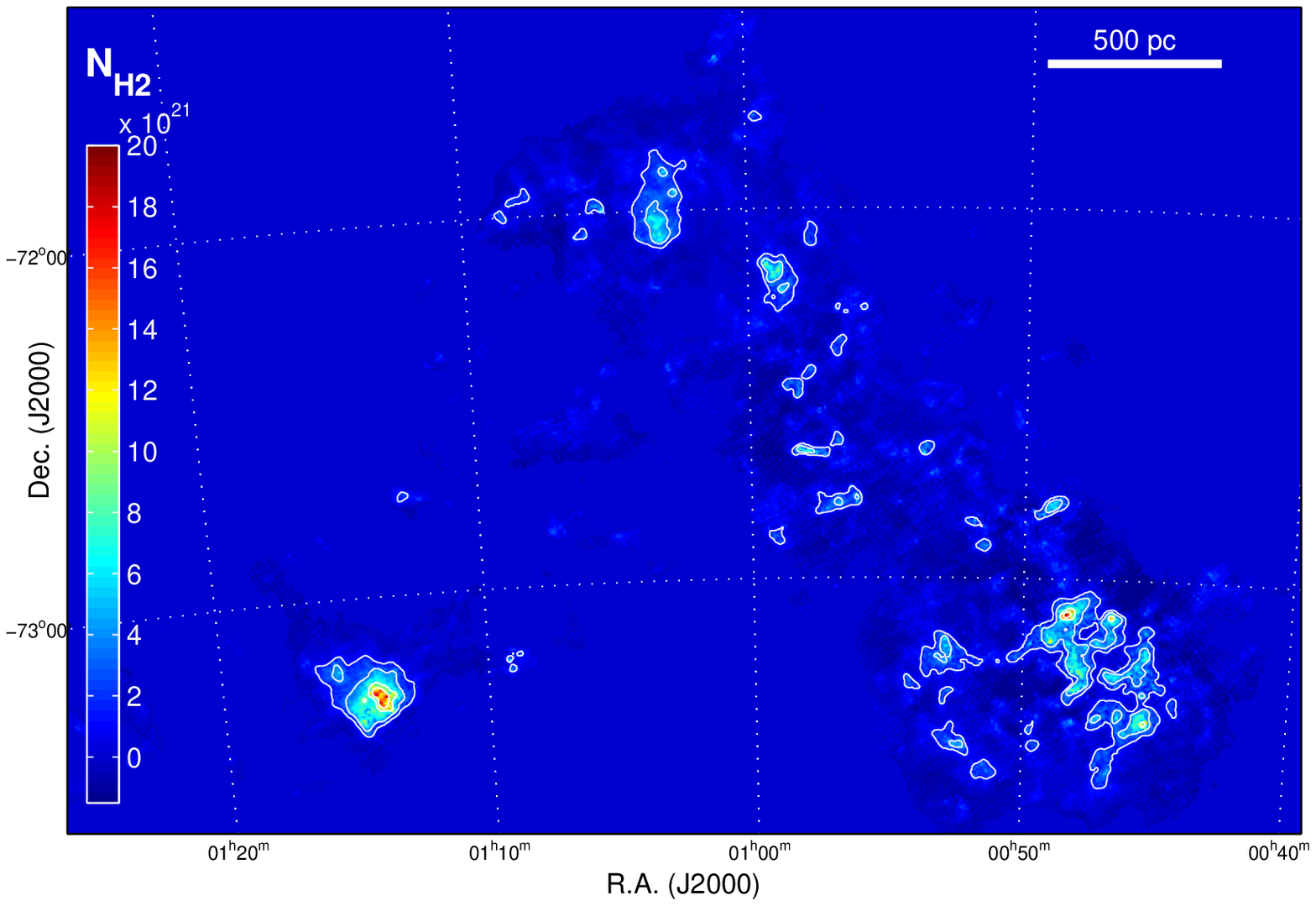}
\plotone{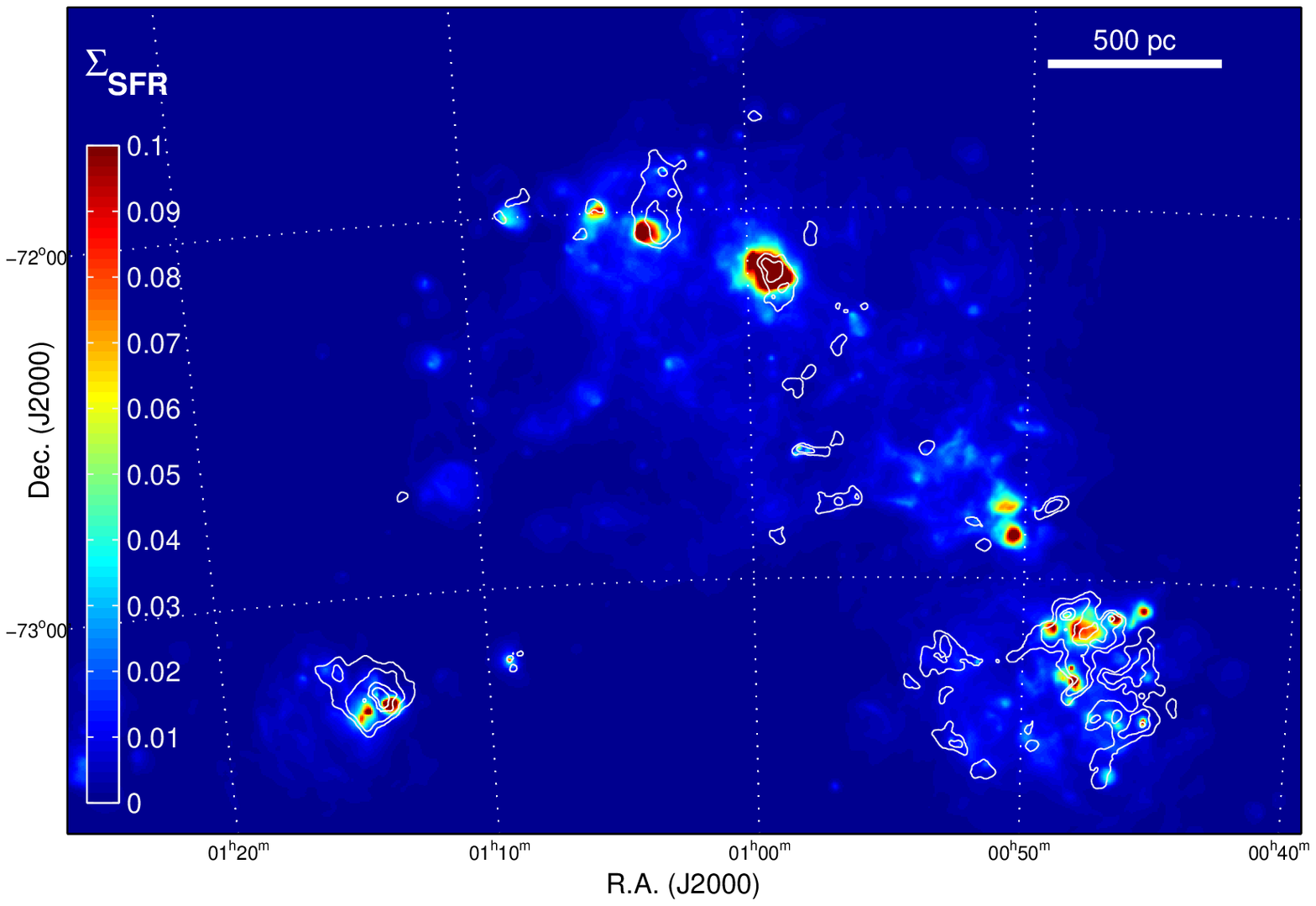}
\epsscale{1}
\caption{{\em (Top)} \htwo\ column density map at $\sim12$~pc resolution. This map
is obtained from modeling the {\em Spitzer} dust continuum
observations from S$^3$MC/SAGE-SMC \citep{BOLATTO07,GORDON11} together
with the combined ATCA/Parkes 21~cm \hi\ map
\citep{STANIMIROVIC99}. 
The colorbar inset indicates the values for the color scale, in units
of $10^{21}$ \percmsq. The N$_{\rm H2}$ contours are placed at N$_{\rm
H2}\approx1.4$,~3.4,~8,~and $12\times10^{21}$ \percmsq, equivalent to
deprojected molecular surface densities $\Smol\approx23, 56,
130,$ and 200~\msunperpcsq\ when including the correction for the cosmic
abundance of helium and the $40^\circ$ inclination of the source. The
western region, oriented roughly SW to NE and harboring most of the
star formation activity as well as the molecular gas is called the SMC
Bar. The extension to the SE is called the SMC Wing, and is
unremarkable in molecular gas except for the N83/N84 molecular cloud
complex, which is the \Shtwo\ peak of the galaxy. {\em (Bottom)}
\htwo\ map overlaid on the unobscured \Ssfr\ map derived from H$\alpha$. The 
SFR is computed using the first term of Eq. \ref{eq:calzetti}, and the
color scale ranges from $\Ssfr=0$ to $0.1$ \msunperyrkpcsq. There is
good correlation between regions with star formation and regions with
molecular gas, but not a one to one correspondence at high spatial
resolution. The overall correlation improves when smoothing to scales
of 200~pc and larger. There is also a pervasive component of diffuse,
low level H$\alpha$ emission.
\label{fig:fig1}
\label{fig:h2map}}
\end{figure*}

\subsection{Estimating H$_2$ from Infrared Dust Emission}
\label{subsec:H2map}

To estimate molecular gas surface densities from far-infrared
emission, we use an extension of the methodology already employed by
\citet{LEROY09}. Here we summarize the main points, but the reader
should refer to that paper for the details, as well as a more thorough
discussion of the effects of several systematics involved in the
derivation of the \htwo\ map. 

We combine \hi\ and infrared (IR) imaging to estimate the distribution
of H$_2$ at a resolution $\theta\sim40\arcsec$ ($r\sim12$~pc).  The IR
data originate from the combination of two {\em Spitzer} surveys,
SAGE-SMC \citep{GORDON11} and S$^3$MC \citep{BOLATTO07}. The \hi\ data
are from \citet{STANIMIROVIC99} and include both interferometric (from
the Australia Telescope Compact Array, ATCA) and single-dish (from
Parkes) observations, and so are sensitive to emission on all spatial
scales. When it is necessary to include an inclination correction we
use $i\approx40^\circ$ for the SMC, as derived from fitting the \hi\
kinematics \citep{STANIMIROVIC04}. Our analysis uses data from {\em
Spitzer}'s MIPS instrument at 70 and 160 $\mu$m. We correct the 160
$\mu$m map for foreground contaminating emission from Galactic cirrus
by subtracting the appropriately scaled \hi\ emission from the Milky
Way, which is identified on the basis of its velocities
\citep[see][]{BOT04,LEROY09}. Our gas surface densities
include a 36\% correction for the mass contribution of helium.

We derive the surface density of \htwo, \Shtwo, by modeling
the IR emission to infer the amount of dust along each line of
sight. Together with a dust-to-gas ratio (or strictly speaking a dust
optical depth to $\hi+\htwo$ surface density ratio), which we estimate
by comparing dust and \hi\ iteratively (see below), the dust emission
allows us to estimate the total amount of gas present. Subtracting the
measured \hi\ surface density yields \Shtwo. We should
caution that this procedure may identify as \htwo\ very cold
self-absorbed \hi\ gas possibly associated with cloud envelopes,
simply because this material will be disproportionally faint at 21~cm
and thus incorrectly subtracted. We include a correction for optical
depth that accounts for self-absorption in the \hi\ map
\citep{DICKEY00}, but such correction is by necessity only
statistical. Thus the procedure we outline may be more accurately
thought of as producing a map of the very cold neutral and molecular
phases, which under normal circumstances would be completely dominated
by the mass of \htwo.

We use the optical depth at 160 $\mu$m, $\tau_{160}$, as our proxy for
the amount of dust along the line of sight. We estimate this quantity
following the method outlined by \citet{LEROY09}, though note that our
calculations improve on theirs in several aspects. We take
the ratio of IR intensities at 100 and 160~$\mu$m ($I_{100}$ and
$I_{160}$) to be a tracer of the equilibrium temperature of a large
grain population that contains most of the dust mass. Because the 100
$\mu$m intensity comes from {\em IRAS}, we only known it at $\sim
4\arcmin$ resolution. To overcome this limitation to the resolution of
the study we compare the $I_{100}/I_{160}$ ratio to the ratio of IR
intensities at 70 and 160~$\mu$m. The emission at 70~$\mu$m, $I_{70}$,
suffers from significant contamination by very small grains out of
equilibrium, and it can only be used to determine the temperature of
the large dust grains after removing the contamination. We calibrate a
relationship that allows us to predict $I_{100}/I_{160}$ from
$I_{70}/I_{160}$, at the coarse resolution of the 100~$\mu$m data.
The fact that the $I_{100}/I_{160}$--$I_{70}/I_{160}$ relation is
well-defined can be appreciated in Figure 2 in \citet{LEROY09}, and
the fitted relations (Equations 2 and 3 in the same paper) suggest
that half of $I_{70}$ is due to emission from stochastically heated
grains. We apply our calibration at the full resolution of the 160
$\mu$m data ($\sim 40\arcsec$) to estimate the dust temperature and
optical depth, $\tau_{160}$. Thus this procedure allows us to take
full advantage of the 160 $\mu$m ($\sim 12$~pc) resolution while
guaranteeing that we match the results derived from the 100 and 160
$\mu$m on size scales larger than 70~pc.

We now have images of $\tau_{160}$, which linearly traces the
distribution of dust, and the \hi\ column density from 21~cm radio
observations corrected for optical depth effects \citep{DICKEY00}. To
create an image of $\Sigma_{\rm H2}$ we combine these maps to produce
local estimates of our gas-to-dust ratio proxy, $\dgdr =
\Sgas / \tau_{160}$. Note that \dgdr\ is an observable in regions
dominated by atomic gas. We use it because the actual value of the
mass emissivity coefficient for dust is unimportant in producing the
\htwo\ map. Its systematic large-scale variations, however, matter and
are very important to consider as we discuss below. Hence

\begin{equation}
\label{eq:h2fromdust}
\Shtwo = \tau_{160}\,\dgdr - \Sigma_{\rm HI}.
\end{equation}

There are several ways to obtain the \dgdr\ from maps of dust and
\hi. One could simply take the average ratio across the entire SMC,
which is certainly \hi\ dominated on large scales. Unfortunately, the
SMC is known to harbor significant large-scale variations in \dgdr,
with higher values in diffuse, low density regions, and particularly
in the SMC Wing \citep{STANIMIROVIC00,BOT04,LEROY07}. Similar large
scale variations, for example, appear to be present in the fraction of
dust in aromatic polycyclic hydrocarbons \citep[PAHs,][]{SANDSTROM10}
and in the optical extinction per gas column density \citep{DOBASHI09}
between lines-of-sight with dense and diffuse gas in the SMC. These
variations are probably not rooted in variations in metallicity. Indeed,
metallicity appears to be approximately uniform across the source
\citep[e.g.,][]{DUFOUR84}, although note that these measurements
are limited to the dense gas phases that form stars. Rather, they may
be due to grain processing in the ISM. These variations could be
analogous to the changes in dust optical depth per unit mass driven by
temperature and grain structure changes observed over smaller scales
in the Milky Way
\citep{BERNARD99,SCHNEE08,FLAGEY09,PLANCKTAURUS11}. An example of a
nearby galaxy which appears to harbor such large-scale variations is
M~31 \citep{NIETEN06}. Evidence suggesting such variations is also
observed in dwarf irregulars, particularly those with extended \hi\
envelopes \citep{WALTER07,DRAINE07}. In the case of the SMC, \dgdr\
appears to be systematically different in the Wing, the SW region of
the Bar, and the NE region of the Bar \citep{LEROY11}.

In the presence of systematic \dgdr\ variations adopting a single
\dgdr, although appealing in its simplicity, will lead to dramatically
overpredicting the amount of \htwo\ in regions that have an
intrinsically lower \dgdr. A more robust approach and considerably
more conservative is to do a local average determination of \dgdr,
either as a function of either position or some other environmental
quantity like IR surface brightness. That is the procedure we follow,
and the \htwo\ results thus obtained are consistent with other
observational constraints, such as virial masses on large scales
\citep{LEROY09} and \htwo\ column densities determined in absorption
along diffuse lines-of-sight \citep{TUMLINSON02}. Nonetheless, \dgdr\
remains the largest source of systematic uncertainty in our estimation
of \Shtwo. It is possible we are missing a large extended molecular
component, although we consider it unlikely.

We implement this approach, making an iterative local measurement of
\dgdr. First, we blank regions of the $\tau_{\rm 160}$
and \hi\ maps where bright CO is detected --- these lines of sight are
likely to include H$_2$, so that $\Shi/\tau_{160}$ is not a good
tracer of the total \dgdr. Next we smooth both maps with a kernel of
radius $R\sim 200$~pc and calculate $\dgdr= \Shi/\tau_{160}$ for each
region at this coarser resolution. This choice of kernel size
represents a compromise between, on the one hand, calibrating \dgdr\
as close as possible to the molecular regions to minimize the effect
of spatial variations, and on the other, avoiding including regions
with significant \htwo\ in its calibration. Using a kernel with
$R\sim450$~pc increases the total molecular mass of the SMC by
$\sim40\%$, while reducing the kernel to $R\sim100$~pc reduces the
mass by $\sim50\%$, thus bracketing the impact of the choice of the
smoothing scale. The \dgdr\ map produced this way has a median value
$\dgdr\approx5.9\times10^{25}$~\percmsq, and a spread of $\sim0.4$ dex.
Representative values for the N83/N84 region near the end of the Wing,
the SW of the Bar, and the NE of the Bar are $\dgdr\sim5.9$, 4.8, and
$3.1\times10^{25}$~\percmsq\ respectively with the highest values
occuring near the central portions of the Wing. These values span the
range of 8 to 14 times higher than Galactic, taking
$\dgdr\approx4.1\times10^{24}$~\percmsq\ determined for Galactic cirrus
as representative of the Milky Way \citep{BOULANGER96}.

We apply this low-resolution \dgdr\ to the nearby blanked regions with
bright CO emission. We then apply Equation \ref{eq:h2fromdust} to our
full-resolution maps to estimate \Shtwo\ everywhere. Because this may
reveal new regions where \htwo\ makes a significant contribution to
the gas surface density we iterate the process once, blanking
everywhere that has bright CO and everywhere with $\Shtwo > 0.5
\Shi$. We verified that iterating further does not significantly
change the molecular surface densities.  The result of this second
iteration is our estimate for \Shtwo\ at an angular resolution
$\theta\sim40\arcsec$ (Figure \ref{fig:h2map}). The map shows very
good correspondence with the recent optical reddening maps from
\citet{HASCHKE11}.

We estimate the uncertainties in this map via a Monte Carlo
calculation. In each iteration of this calculation, we add realistic
noise to the data, adjust the zero point of the IR maps within the
uncertainties, rederive the relation used to predict $I_{70}/I_{160}$
from $I_{100}/I_{160}$ and propagate the noise in this relation
forward. We also vary the wavelength dependence of the dust opacity,
$\beta$, across its likely range $\beta\approx1-2$, with one value
randomly chosen for each iteration. We carry out 1000 such exercises
and calculate the 1, 2, and 3$\sigma$ uncertainties in $T_{\rm dust}$
and $\tau_{160}$ from the resulting distributions. These are
propagated into uncertainties in $\Sigma_{\rm H2}$. Using this
technique we estimate our uncertainty in $\Shtwo$ before deprojection
to be $1\sigma\sim15$ \msunperpcsq\ (equivalent to $\Shtwo\sim11.5$
\msunperpcsq\ after correcting for the inclination angle of the SMC).

The following points are important to keep in mind, since they
represent limitations of our \htwo\ map and ultimately of our
analysis. First, to remove spurious \htwo\ emission from our map we
only retain contiguous regions of area $\geq4$ square arcminutes (this
cleans up a few islets of emission) and $\Shtwo>23$
\msunperpcsq\ ($2\sigma$ deprojected), and set the rest of the map to
$\Shtwo = 0$. Because of this and the method used to derive a local
\dgdr, we cannot recover a pervasive
\htwo\ component. Other observations suggest that a large molecular
component of this form is very unlikely \citep{DICKEY00,TUMLINSON02},
thus we do not view this as a significant uncertainty although it
remains a possibility.

Second, the SMC has a complex geometry (see discussion in
\S\ref{subsec:smcintro}). As a result certain lines of sight may
contain a combination of regions with different
\dgdr. The presence of diffuse, high-\dgdr\
emission along the line of sight will invalidate Equation
\ref{eq:h2fromdust}. The easiest way to correct for this is to
subtract a ``dust-free'' component from the \hi\ map to adjust the
zero point of the dust--gas correlation. The likely magnitude of this
dust-free \hi\ component for the SMC is $\Shi\sim20-40$
\msunperpcsq\ \citep{LEROY11}. We do not remove a ``dust-free'' 
component in the analysis presented here. Removing it would drive
our \htwo\ map toward lower values of \Shtwo.

Third, despite our attempt at minimizing the effect of systematic
\dgdr\ changes through a local determination, variations in the 
dust emissivity and \dgdr\ between the dense and diffuse ISM are
possible and largely unconstrained. If they have the expected sense of
higher optical depths per unit gas in \htwo\ than in \hi, they will
drive the \Shtwo\ toward values that are too high. Thus the errors
introduced have the same sign as those discussed in the previous
point. Our best estimate of their combined effect is a factor of two
systematic uncertaintly in \Shtwo, with lower \Shtwo\ values more
likely.

Fourth, at 98\arcsec\ the resolution of the \hi\ map is somewhat lower
than that of the IR maps. Therefore at spatial resolutions better than
$r\sim 30$~pc we have assumed that \Shi\ is smooth at scales below the
$1.6\arcmin$ resolution of our \hi\ map \citep{STANIMIROVIC99}. This
assumption is almost certainly flawed in detail, but will average out
across the whole galaxy, and it has no impact on quantities determined
on $r\gtrsim30$~pc scales.

Our estimate of a factor of 2 systematic uncertainty in our
\htwo\ determination, together with source geometry uncertainties 
discussed in \S\ref{subsec:smcintro} and \S\ref{subsec:gassfr}, result
in a combined systematic uncertainty of $\sim0.5$ dex (a factor of
3) in the molecular depletion time in the SMC (\S\ref{subsec:molsfr}).

\subsection{Estimating H$_2$ from CO}
\label{subsec:COmap}

The standard practice is to estimate the amount of molecular gas
present in a system using $^{12}$CO observations. This requires the 
use of a proportionality factor to convert intensity into column
density or mass. We use the following equations

\begin{eqnarray}
N(\htwo) &=& \xco\,\Ico \\
\Mmol  &=& \aco\,L_{\rm CO},
\end{eqnarray} 

\noindent where our adopted proportionality constants appropriate for
Galactic gas are $\xco=2\times10^{20}$ \xcounits\ and
$\aco=4.4$ \acounits, \Ico\ is the integrated
intensity of the $^{12}$CO \jone\ transition (usually in \Kkmpers),
and L$_{\rm CO}$ is the luminosity of the same transition (in
\Kkmperspcsq).  The resulting column density and mass are in \percmsq\
and M$_\odot$ respectively, and the molecular mass \Mmol\ corresponds
to the mass of \htwo\ corrected by the contribution of the cosmic
abundance of He.  Note that although the Galactic values are
approximately appropriate on the small spatial scales in the CO-bright
material \citep{BOLATTO08}, they are most likely inappropriate on the
large scales \citep{RUBIO93b,ISRAEL97,ISRAEL03,LEROY07,LEROY11}. This
is not surprising: if CO and \htwo\ are not perfectly co-located their
ratio will depend on the regions over which we are averaging. In
particular, if CO is confined to the highly shielded high surface
density cores while \htwo\ is more widespread
\citep[e.g.,][]{LEQUEUX94,BOLATTO99}, \xco\ will be small on CO-bright
regions and large when measured on large spatial scales (CO freeze-out
into grains is not important on the spatial scales considered).  We
discuss this further in \S\ref{sec:results}.

\subsection{Tracing Recent Star Formation}
\label{subsec:SFRmap}

We use \ha\ emission, locally corrected for extinction, to trace the
surface density of recent star formation. The
\ha\ observations we use were obtained, calibrated, continuum
subtracted, and mosaiced by the Magellanic Cloud Emission Line Survey
\citep[MCELS,][]{SMITH99}.
Particularly for bright stars, the continuum subtraction may leave
noticeable artifacts, which we have masked out in our analysis. We
convolve these images with the point spread function of the MIPS 160
$\mu$m camera, to match them to the $\sim40\arcsec$ resolution of our
molecular gas map.

We correct the ionizing photon flux inferred from \ha\ for the effects
of extinction using the MIPS 24 $\mu$m image obtained by SMC-SAGE, and
the prescription from \citet[][]{CALZETTI07}. The implied extinctions
are very small over most of the SMC, only becoming significant for the
centers of the brightest \hii\ regions. This is consistent with the
findings of
\citet{CAPLAN96}, who used the Balmer decrement technique to
obtain a typical extinction $A_{\rm H\alpha} \sim0.3$~mag toward
bright SMC \hii\ regions \citep[see also][for a discussion of
extinction based on fitting color--magnitude diagrams]{HARRIS04}. The
small extinction correction should not be surprising since the SMC is
notoriously dust poor, with a dust-to-gas ratio that is a factor of
$5-10$ lower than Galactic
\citep{LEROY07}. This galaxy simply does not harbor much obscured star
formation activity.

The combination of \ha\ and 24$\mu$m maps yields an estimate of the
extinction-corrected ionizing photon surface density. For our
analysis we convert this to a star formation rate surface density,
$\Sigma_{\rm SFR}$, following 
\citep[][Equation 7]{CALZETTI07}, 

\begin{equation}
\Ssfr = 5.3\times10^{-42} \left[ \Sigma_{\rm H\alpha} + 0.031\Sigma_{24}\right] \label{eq:calzetti}
\end{equation}

\noindent where $\Sigma_{\rm H\alpha}$ and $\Sigma_{\rm 24}$ are
respectively the surface densities of \ha\ and 24 $\mu$m far-infrared
emission, in erg~s$^{-1}$~kpc$^{-2}$, and the relation to SFR assumes
an underlying broken power-law Kroupa initial mass function.  This
prescription is local, obtained for individual star forming regions
and calibrated against Paschen $\alpha$ emission. The use of 24$\mu$m
emission to correct \ha\ has been tested in M~33 by \citet{RELANO10}
on the scale of individual \hii\ regions. \citet{CALZETTI07} found the
metallicity dependence of the extinction correction to be $\sim20\%$
for galaxies down to nebular metallicities below that of the SMC.

The very high spatial resolution of our data, $r\sim 12$~pc (though
much longer along the line of sight) adds some complication to the
concept of a star formation rate. The calibration by
\citet{CALZETTI07} is 
derived on spatial scales of hundreds of parsecs. On the small scales
corresponding to our full resolution any individual line-of-sight may
poorly sample the high mass end of the initial mass function, be
populated by stars resulting from a single star formation episode, or
emit \ha\ radiation resulting from ionizing photons originating in an adjacent
region. Spatial smoothing to larger scales removes these concerns.
Throughout this paper we consider spatial scales ranging from $r\sim
12$~pc to the whole SMC. At the smallest of these scales, the idea of
local star formation rate may break down, but at resolutions of 200~pc
or 1~kpc (and certainly for the whole SMC) the application of a
standard H$\alpha$ to SFR conversion should be adequate. These scales
also allow a fairer comparison to most other extragalactic studies.
  
It is important to note that in the results presented in the following
sections we do not remove a diffuse ionized gas component, which may
be as high as $40\%$ of the total \ha\ emission in the SMC
\citep{KENNICUTT95}. The $200-600$~pc scales on which \citet{CALZETTI07} 
performed their calibration are much larger than the $\sim12$~pc on
which we carry out our study. Thus it seems likely that part of the
diffuse ionized emission due to the escape of ionizing photons from
\hii\ regions is already included in the calibration, suggesting that
$40\%$ is an upper bound to the SFR correction due to diffuse
emission.  To assess the impact of the diffuse \ha\ on our results we
performed an analysis where we spatially filter the \ha\ image on
several scales to remove the diffuse component, along the lines
described in \citet[][and references therein]{RAHMAN11}.  We found
that the results presented in the following sections are robust to the
presence of diffuse \ha. The main effect of removing a diffuse ionized
component is to correspondingly lengthen the gas depletion time, or
equivalently somewhat lower the local star formation efficiency.

\section{Results and Discussion}
\label{sec:results}

Maps of \Shtwo, \Sgas, and \Ssfr\ allow us to study several aspects of
the star formation law. In the following sections we will characterize
the global properties of the SMC, compare molecular gas and star
formation (the ``molecular star formation law''), total neutral gas
and star formation (the ``total gas star formation law''), and measure
the molecular fraction as a function of surface density.

\subsection{Global Properties}
\label{subsec:global}

Our study allows us to derive a number of interesting integrated
properties for the SMC. Integrating our extinction-corrected \ha\ map,
we obtain a global star formation rate ${\rm SFR_{SMC}}\sim0.037$
\msunperyr .  The extinction correction accounts for $\sim10\%$
of the global SFR. About 30\% of the integrated \ha\ emission arises
from extended low surface brightness regions, with equivalent star
formation rate densities of $\Sha<5\times10^{-3}$ \msunperyrkpcsq,
likely reflecting long mean free paths for ionizing photons escaping
\hii\ regions (with perhaps some contribution from other sources
of ionization). The global SFR we measure is very comparable to the
present SFR obtained from the photometric analysis of the resolved
stellar populations by \citet{HARRIS04}, and only slightly lower than
the ${\rm SFR_{SMC}}\sim0.05$ \msunperyr\ obtained by
\citet{WILKE04} based on the study of the SMC far-infrared
emission. Note, however, that in the latter case the estimate is based
on correcting up considerably the standard FIR calibration by assuming
a much smaller dust optical depth in the SMC. It also relies on the
synthetic model grid of \citet{LEITHERER95}, which has been superseded
by the newer models incorporated in the \citet{CALZETTI07}
calibration. In any case, this range of values is representative of
the typical uncertainties in determining SFRs, also applicable to our
value.

Integrating the molecular mass over the regions of the map with signal
above our 2$\sigma$ threshold of 23 \msunperpcsq , we obtain ${\rm
  M_{SMC}^{mol}}\sim2.2\times10^7$ \msun . This is a $\sim35$\%
revision down from previous studies using the same technique
\citep{LEROY07}, mostly due to differences in the estimation of the
dust-to-gas ratio and well within the factor of $\sim2$ systematic
uncertainties present in the analysis \citep[for discussions
see][]{LEROY07,LEROY09}. The global CO luminosity of the SMC is
approximately $1\times10^5$ \Lcounits\ \citep[][assuming a 20\%
correction for flux in unmapped regions]{MIZUNO01}. This yields a
globally averaged conversion factor from CO to molecular mass of
$\aco\sim220$ \acounits. We return to \aco\ in the next section.

The gas depletion time, $\tau_{\rm dep}$, is a convenient way to
parameterize the normalized star formation rate: $\tau_{\rm dep}$ is the
time needed for the present SFR to exhaust the existing gas reservoir
\citep{YOUNG86}. Sometimes the inverse of this quantity is presented
as a so-called star formation efficiency (\SFE), indicating the
fraction of the gas reservoir used in $10^8$ yr. Galaxies with high
\SFE\ can only sustain their present star formation activity for a
short period of time, and thus are experiencing a starburst.

From this integrated SFR and \Mmol, the global molecular gas depletion
time of the SMC is approximately $\tau_{\rm dep}^{\rm mol} = \Mmol/SFR
\sim0.6$ Gyr. For comparison, the global molecular depletion time
inferred from the CO luminosity and a Galactic conversion factor
($\alpha_{\rm CO} = 4.4$) would be $\sim0.01$~Gyr. Although a Galactic
conversion factor is clearly inappropriate, this exercise highlights
the need for an alternative tracer of H$_2$. The CO in the SMC is
either dramatically underluminous for the observed level of star
formation activity, or the SMC is in the midst of a starburst that
will exhaust the molecular gas reservoir in only $\sim
10^{7}$~yr. Although the study of the global star formation history of
the SMC by \citet{HARRIS04} finds its present SFR to be somewhat
larger than the past average, the magnitude of that effect cannot
explain such an implausibly short molecular gas depletion time.

Given its atomic mass \citep[$M_{\rm
HI}\simeq4.2\times10^8$~M$_\odot$;][]{STANIMIROVIC99}, the SMC keeps
only $\sim 5\%$ of its gas in the molecular phase. Thus it appears
genuinely poor at forming GMCs. The corresponding total gas
depletion time is $\tau_{\rm dep}^{\rm gas}\sim11.8$~Gyr, so the SMC has
enough fuel to sustain its current rate of star formation for
approximately a Hubble time.
By comparison, the median total gas depletion time for the 12 large
galaxies in the THINGS/HERACLES sample is $\tau_{\rm dep}^{\rm
gas}\approx6.0$~Gyr \citep{LEROY08}.  Although a wide range of
depletion times are present, $1<\tau_{\rm dep}^{\rm gas}<14$ Gyr, 10
of these galaxies have $\tau_{\rm dep}^{\rm gas}<8$~Gyr. In contrast,
in \hi-dominated dwarf galaxies and the outer disks of spirals $\tau_{\rm
dep}^{\rm gas}\gtrsim10$~Gyr \citep{BIGIEL10b}.

\subsection{Relation Between Molecular Gas and Star Formation}
\label{subsec:molsfr}

\begin{figure}[h!]
\plotone{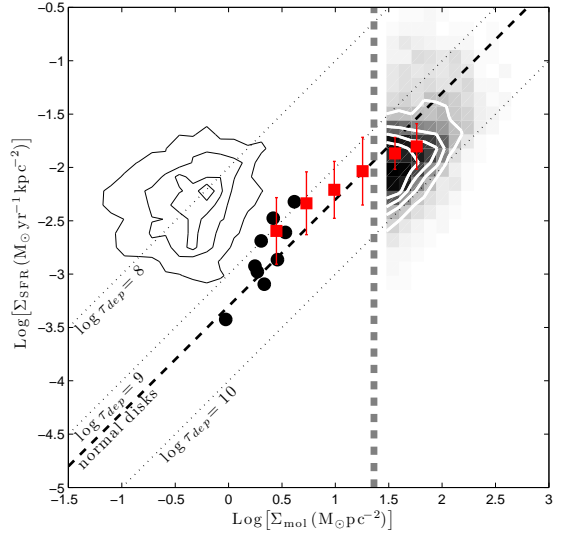}
\caption{Molecular star formation law in the SMC. The gray scale shows
the binned two-dimensional distribution of the \Ssfr\ to \Smol\
correlation at a resolution $r\sim12$~pc, where \Smol\ is derived from
the dust modeling. The intensity scale is proportional to the number
of points that fall in a bin, with white contours indicating levels
that are 20\%, 40\%, 60\%, and 80\% of the maximum. The vertical gray
dashed line indicates our adopted 2$\sigma$ sensitivity cut for the
$r\sim12$~pc data. The red squares and black circles show the results
after spatial smoothing to $r\sim200$~pc and $r\sim1$~kpc,
respectively (the sensitivity limit has been moved down accordingly,
assuming Gaussian statistics). The bars in the $r\sim200$~pc data show
the standard deviation after averaging in \Smol\ bins.  The black
contours, placed at the same levels as the white contours, show the
distribution of \Smol\ derived from CO observations \citep{MIZUNO01}
using the Galactic CO-to-\htwo\ conversion factor. The dotted lines
indicate constant molecular gas depletion times $\tau_{\rm dep}^{\rm
mol}=0.1$, 1, and 10 Gyr. The dashed line indicates the typical
depletion time for normal disk galaxies $\tau_{\rm dep}^{\rm
mol}\sim2\times10^9$ years
\citep{BIGIEL08}. The $\tau_{\rm dep}^{\rm mol}$ 
in the SMC is consistent with the range observed in normal disks for
\Smol\ derived from dust modeling. \label{fig:fig2}\label{fig:sflmol}}
\end{figure}

Figure \ref{fig:sflmol} shows the correlation between molecular gas
and star formation activity traced by the extinction-corrected \ha\ at
several spatial resolutions: $r\sim12$~pc (grayscale), $r\sim200$~pc
(red squares, binned), and $r\sim1$~kpc (black circles). The gray area
and the corresponding white contours show that a fairly
well-defined relation exists between the star formation activity and the
surface density of molecular gas obtained from the dust continuum
modeling described in \S\ref{subsec:H2map} even on our smallest
scales.

Note that the $x$ and $y$ axes of this plot are not implicitly
correlated. The abscissa contains information from the 70, 100, and
160~$\mu$m far-infrared continuum as well as the \hi\ map, while the
ordinate is chiefly \ha. The small extinction correction derived from
the 24 $\mu$m data has a negligible effect on the correlation. Also
note that because we correct for dust temperature when deriving the
dust surface density $\Smol$ should, in principle, also be independent
of heating effects.

The molecular gas depletion time depends on the scale considered
(Figure \ref{fig:sflmol}). On the smallest scales considered,
$r\sim12$~pc, the depletion time in the molecular gas is
$\log[\tau_{\rm dep}^{\rm mol}/(1 {\rm Gyr})]\sim0.9\pm0.6$
($\tau_{\rm dep}^{\rm mol}\sim7.5$ Gyr with a factor of 3.5
uncertainty after accounting for observed scatter and systematics
involved in producing the \htwo\ map as well as the geometry of
the source). As mentioned in the previous
section, $\tau_{\rm dep}^{\rm mol}$ shortens when considering larger
spatial scales due to the fact that the \ha\ and \htwo\ distributions
differ in detail, but are well correlated on scales of hundreds of
parsecs (Fig. \ref{fig:fig1}). On size scales of $r\sim200$~pc (red
squares in Figure \ref{fig:sflmol}), corresponding to very good
resolution for most studies of galaxies beyond the Local Group, the
molecular depletion time is $\log[\tau_{\rm dep}^{\rm mol}/(1 {\rm
Gyr})]\sim0.2\pm0.3$, or $\tau_{\rm dep}^{\rm mol}\approx1.6$~Gyr. The
depletion time on $r\sim1$~kpc scales (black circles in Figure
\ref{fig:sflmol}), corresponding to the typical resolution of
extragalactic studies, stays constant for the central regions of our
map (where the smoothing can be accurately performed), $\log[\tau_{\rm
dep}^{\rm mol}/(1 {\rm Gyr})]\sim0.2\pm0.2$. Thus our results converge
on the scales typically probed by extragalactic studies. This
constancy reflects the spatial scales over which \ha\ and molecular
gas are well correlated. Although the precise values differ, a very
similar trend for $\tau_{\rm dep}^{\rm mol}$ as a function of spatial
scale is observed in M~33 \citep{SCHRUBA10}. The further reduction of
the depletion time when considering the entire galaxy ($\tau_{\rm
dep}^{\rm mol}\approx0.6$~Gyr) reflects the contribution from a
component of extended \ha\ emission, which is filtered out in the
calibration of the SFR indicator \citep{CALZETTI07,RAHMAN11}. The SMC
is on the high end of the observed distribution of values for the
fraction of diffuse \ha, but fractions of $40\%-50\%$ are common in
galaxies \citep[e.g.,][]{HOOPES99}. 

Within the uncertainties, our results are not significantly different
from the mean \htwo\ depletion time obtained in studies of
molecule-rich late-type disks on 750~pc to 1~kpc spatial scales, where
$\tau_{\rm dep}^{\rm mol}\sim2.0\pm0.8$ Gyr averaged over regions with
molecular emission \citep[$\SFE\approx5\%$ within 0.1
Gyr,][]{BIGIEL08,LEROY08,BIGIEL11}. This is the methodology used in
resolved studies of $\tau_{\rm dep}^{\rm mol}$ in more distant
galaxies, and our results are directly comparable.

This contrasts sharply with the results using the CO map (black
contours) obtained by the NANTEN telescope \citep[][]{MIZUNO01}, using
a Galactic CO-to-\htwo\ conversion (the assumption in many
extragalactic studies). The CO distribution is offset by a factor of
$\sim40$ from the \htwo\ distribution. This offset corresponds to the
most common \aco\ implied by our dust-map, $\aco\approx185$ \acounits
(very similar to the global \aco ). Both values of \aco\ are broadly
consistent with CO-to-\htwo\ conversion factors obtained by previous
dust continuum modeling and virial mass techniques on large scales
\citep{RUBIO93b,ISRAEL97,LEROY07,LEROY11}, though factors of 2--3
discrepancy persist from study-to-study. They differ, however, from
estimates based on small-scale virial masses toward the CO-bright
peaks, which tend to obtain values of \aco\ closer to Galactic
\citep{ISRAEL03,BLITZ07,BOLATTO08,MULLER10}. This discrepancy between
CO-to-\htwo\ conversion factor on the large and on the small scales
can be understood in terms of the existence and mass dominance of
large molecular envelopes poor in CO. Such envelopes are expected at
all metallicities \citep{GLOVER10}, and at the low metallicity of the
SMC they likely constitute the dominant reservoir of molecular gas
\citep{WOLFIRE10}.

\subsection{Relation Between Total Gas and Star Formation}
\label{subsec:gassfr}

\begin{figure}[t]
\plotone{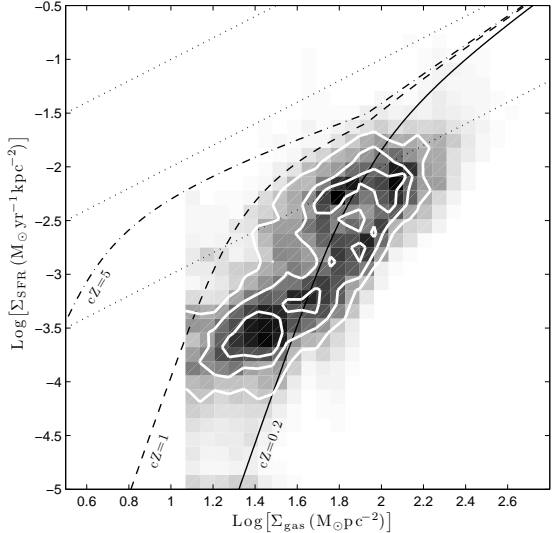}
\caption{Total gas star formation law in the SMC. The gray scale shows
the two dimensional distribution of the correlation between \Ssfr\ and
\Sgas, where \Sgas\ is the surface density of atomic plus molecular
gas corrected by Helium. The white contours indicate the correlation
due to atomic gas alone, which dominates the gas mass (and \Sgas) in
the SMC. The contour levels, and the dotted lines indicating constant
$\tau_{\rm dep}^{\rm gas}$, are at the same values as in Figure
\protect\ref{fig:sflmol}. The dash-dotted, dashed, and solid lines
indicate the loci of the model by
\citet{KRUMHOLZ09} (KMT09) for clumping-factor by metallicity products
$cZ=5,$ 1, and 0.2 respectively. The first two bracket the behavior of
most galaxies observed at 750~pc resolution (see KMT09, Figure 1),
while $cZ=0.2$ would be the value expected for the SMC with unity
clumping-factor (a reasonable assumption for the spatial resolution of
the observations presented here, $r\sim12$~pc). Note that the surface
density at which \hi\ starts to saturate in the SMC is $\Shi\sim50$
\msunperpcsq\ (the typical surface density is $\Shi\sim85$ \msunperpcsq\ at 
the full resolution of the \hi\ data), considerably larger that the
typical value in normal metallicity galaxies where $\Shi\lesssim10$
\msunperpcsq\
\citep{BIGIEL08,BIGIEL11}. As a consequence any use of a ``standard'' total
gas-star formation correlation for the SMC would dramatically
underpredict total gas surface densities, or overpredict star
formation activity. This is not true for molecular gas, as we discuss
in the previous figure. \label{fig:fig3}\label{fig:sflgas}}
\end{figure}

Figure \ref{fig:sflgas} shows the total gas star formation law for the
SMC, the relation between \Ssfr\ and total (\hi\ + H$_2$) gas surface
density 
\Sgas, as well as \Ssfr\ vs. \Shi\ (which is almost the same, as atomic gas 
dominates).
This relationship may be more complex than the molecular star
formation law, resulting from a combination of phase balance in the
ISM and the relative efficiencies of different types of gas at forming
stars. Using a power-law ordinary least-squares bisector fit we find that
$\log(\Ssfr)=(2.2\pm0.1)\log(\Sgas)+(-6.5\pm0.1)$ for $\Sgas>10$
\msunperpcsq\ at the full spatial resolution of the observations.
This is very similar to the typical $1+p\approx2$ slope measured for
this relation in \hi-dominated regions of galaxies \citep{BIGIEL10b},
or the typical power-law index of the \Ssfr\ to \Shi\ relation observed
in faint dwarfs by \citet{ROYCHOWDHURY09}.

Thus the relation between \Ssfr\ and \Sgas\ is steep in the SMC, and
similar to \Ssfr\ vs. \Sgas\ at low surface density in normal
galaxies, but the relation in the SMC is noticeably displaced toward
larger total gas surface densities compared to large spirals
\citep[this was already evident in the study by][]{KENNICUTT95}. The
atomic surface density in normal metallicity galaxies almost never
exceeds a saturation point of $\Shi\approx10$
\msunperpcsq\ \citep{BIGIEL08} 
averaged over $\sim$ kpc scales, although smaller-scale higher-column 
\hi\ clouds are observed in the Milky Way \citep[e.g.,][]{HEILES03}.
That is not the case in the SMC, where \Shi\ reaches values of $\sim100$
\msunperpcsq\ ($N(\hi)\sim1\times10^{22}$ \percmsq, see the white
contours in Figure \ref{fig:fig3}). This is not purely the result of
the high spatial resolution; high \Shi\ persists even averaged over
large spatial scales \citep{STANIMIROVIC99}. It may be partially
enhanced by the complex geometry of the source; the SMC is an
interacting galaxy that may have significant elongation along the line
of sight, not a simple flat disk \citep[e.g.,][]{YOSHIZAWA03}. All
surface densities (averaged over large scales) would be reduced by a
factor $\cos\,i/0.77$ if the geometry is disk-like and inclination
exceeds 40$^\circ$, or possibly by a larger factor if the galaxy is
also elongated along the line of sight. These corrections are not
large enough to explain the full magnitude of the effect
we see, in light of the \hi\ kinematic analysis by
\citet{STANIMIROVIC04}, but could conceivably contribute a factor
of $\sim1.5-2$ (where the 50\% factor corresponds to changing the
inclination from $i=40^\circ$ to $i=60^\circ$, and a further 20\%-30\%
factor is an estimate for elongation, obtained by evaluating the
contribution from gas at high velocities to \Shi, see
\S\ref{subsection:molgasfraction}).

This displacement of appreciable star formation activity towards high
gas surface densities means that applying the Kennicutt--Schmidt
relation observed for normal galaxies to the total gas surface density
in the SMC would lead to a dramatic overprediction of the star
formation rate.  For example, in a normal disk we would expect total
gas surface densities of $\Sgas\sim20$ \msunperpcsq\ to be associated
with star formation rates of $\Ssfr\sim0.01$ \msunperyrkpcsq\ on
$\sim1$~kpc scales \citep{BIGIEL08,BIGIEL11}, while in the SMC the
typical SFR associated with such gas surface density is one to two
orders of magnitude lower. Conversely, using the observed SFR to
estimate gas content (as it is sometimes done in studies of the high
redshift universe) would lead to dramatic underestimates of the amount
of {\em total gas} present.

\begin{figure}[h]
\plotone{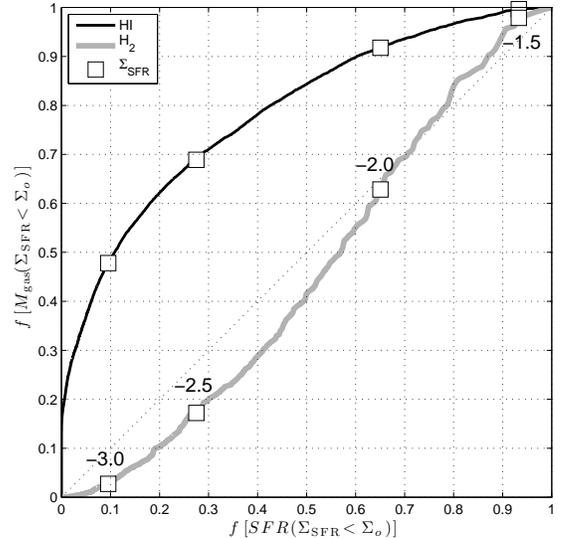}
\caption{Cumulative distribution function of gas mass in lines-of-sight 
with increasing star formation activity at a spatial resolution of
200~pc, plotted against the cumulative distribution function of SFR
for the same lines-of-sight. The abcissa corresponds to the fraction
of the total SFR below a particular value of the \Ssfr. The ordinate
corresponds to the fraction of gas (\hi\ in black, \htwo\ in gray)
accumulated in those lines of sight. The squares show the locations of
particular values for $\log[\Ssfr]$, in units of \msunperyrkpcsq. For
example, about 28\% of all the extinction-corrected \ha\ emission in
the SMC is found in lines-of-sight with corresponding surface
densities $\Ssfr\lesssim10^{-2.5}$
\msunperyrkpcsq, and those lines-of-sight contain $\sim69\%$ and
$\sim17\%$ of all the atomic and molecular mass, respectively. This
plot highlights the fact that most \hi\ is found in regions with
little star formation, even on scales of 200~pc. \label{fig:fig4}\label{fig:cumdist}}
\end{figure}

This offset in the total gas star formation law stands in stark
contrast to the conclusions in
\S\ref{subsec:molsfr} about the relation between star formation
activity and molecular gas, where the SMC appeared very similar to
normal disk galaxies. The fact that the molecular \SFE\ in the SMC
is within the range observed in other galaxies implies that, within
the usual (factor of 3) uncertainties, the observed star formation
rate accurately reflects the amount of molecular gas present.

This difference implies that there are more quantitative similarities
between the distributions of molecular gas and recent star formation
than between atomic gas and recent star formation. We plot this
directly in Figure \ref{fig:cumdist}, which compares the cumulative
distributions of star formation and gas surface density at $\sim
200$~pc resolution, a few times larger than the size of a large GMC in
the Milky Way. This represents a typical size scale over which gas and
star formation should be correlated.  The abcissa in Figure
\ref{fig:cumdist} corresponds to the fraction of the total SFR below a
particular value of the \Ssfr. The ordinate corresponds to the
fraction of gas accumulated in those lines of sight.  

Figure \ref{fig:cumdist} shows that star formation and molecular gas
track linearly with each other, while the nonlinear shape of the \Shi\
distribution reflects the fact that most of the \hi\ is found on lines
of sight with little star formation activity. A similar phenomenon is
observed in other faint dwarf galaxies
\citep{ROYCHOWDHURY09,ROYCHOWDHURY11}.  This lack of correspondence
between \hi\ and \Ssfr\ is even more accute at $r\sim12$~pc, while
\htwo\ and star formation continue to track each other
well. Approximately 85\% by mass of the
\hi\ in the SMC is in the warm phase, according to observation
and modeling of 28 lines-of-sight toward background continuum sources
\citep{DICKEY00}. Together with Figure \ref{fig:cumdist}, this
strongly suggests that most \hi\ in the SMC is not directly related to
the star forming gas, i.e., it does not belong to atomic envelopes of
molecular clouds.

\subsection{Molecular Gas Fraction}
\label{subsection:molgasfraction}

\begin{figure}[h]
\plotone{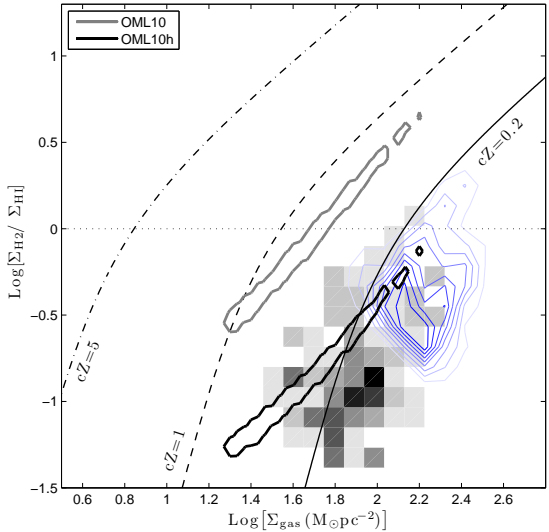}
\caption{Ratio of molecular to atomic gas in the SMC. The blue contours 
and the gray scale show the two-dimensional distribution of the ratio
\Shtwo/\Shi\ versus \Sgas\ on scales of $r\sim12$~pc and $r\sim200$~pc
respectively (note that the hard edge present in the blue contours at
low ratios and low \Sgas\ is the result of our adopted $2\sigma$ cut in
\Shtwo).
The dotted horizontal line indicates $\Shtwo/\Shi=1$, denoting the
transition between the regimes dominated by \hi\ and
\htwo. The dash-dotted, dashed, and solid lines show the predictions
of KMT09 for different values of the $cZ$ parameter, as in Figure
\ref{fig:sflgas}. For the SMC, $cZ=0.2$ at $r\sim12$~pc, and the KMT09
curve overestimates the molecular to atomic ratio by a factor of $2-3$. 
At $r\sim200$~pc we may expect $cZ\approx1$ using the standard clumping
factor $c=5$ adopted by KMT09 for unresolved complexes.  Although
molecular gas in the SMC is highly clumped, the atomic gas is not, so
the $cZ=1$ curve overestimates
\Shtwo/\Shi\ at 200~pc.  The thick gray and black contours indicate the
predicted surface density ratio of gas in gravitationally bound
complexes to diffuse gas, $\Sigma_{\rm gbc}/\Sigma_{\rm diff}$, in
OML10 and in the model modified to incorporate the extra heating of in
the diffuse gas (OML10h, see \S\ref{subsec:heating}),
respectively. The contours are calculated for the metallicity and
distribution of stellar plus dark matter density in the SMC. The
original OML10 prediction for $\Sigma_{\rm gbc}/\Sigma_{\rm diff}$ is
considerably higher than the observed
\Shtwo/\Shi.  The tightness of the contours is due to the fact that
the self-gravity of the gas dominates over the stellar plus dark
matter contribution, thus there is an almost one-to-one correspondence
between \Sgas\ and the prediction for $\Sigma_{\rm gbc}/\Sigma_{\rm
diff}$.
\label{fig:fig5}\label{fig:rH2}}
\end{figure}

Figure \ref{fig:rH2} shows our final basic observational result, the
molecular fraction of the SMC as a function of total gas surface
density. Grayscale shows the density of our data in $(\Sigma_{\rm
H2}/\Sigma_{\rm HI})$--$(\Sigma_{\rm gas})$ space at 200~pc resolution,
the blue contours show it at 12~pc resolution. The curved labelled
lines indicate model predictions by
\citet{KRUMHOLZ09}. At $\sim$ kpc resolution,
most massive star-forming disk galaxies lie between the curves
labelled $cZ=1$ and $cZ=5$ (see \S\ref{subsec:kmt09}). For a given
total gas surface density, the SMC has molecular fractions much lower
than these large galaxies with the offset often an order of magnitude
or more. We discussed in \S\ref{subsec:smcintro} the fact that the
complex distribution of \hi\ in the SMC along the line-of-sight is a
source of uncertainty. We can obtain a rough estimate of the effects of
\hi\ not associated with the disk of the SMC on the molecular ratio
by recalculating \Shi\ after removing \hi\ emission outside the
velocity range $v_{lsr}\approx120-180$~\kmpers, taken to be
representative of the disk of the galaxy. We find this exercise lowers
\Shi\ by at most 30\% the faint regions of the Wing, and more
typically $\sim10\%-15\%$. This is a small correction in the molecular
ratio, well within our uncertainties in \Shtwo\ alone, and although it
may significantly contribute to the observed dispersion in
$\Shtwo/\Shi$ it cannot be the cause of its offset with respect to
normal galaxies. The SMC is strikingly bad at turning its wealth of
atomic gas into molecular gas, particularly given the very high
surface densities found in this galaxy.

\subsection{Synthesis of Results}

Using our dust-based \Shtwo\ map, we showed that to first order the
molecular star formation law in the SMC resembles that in disk
galaxies. There is still room within the uncertainties for a factor of
2--3 decrease in $\tau_{\rm dep}^{\rm mol}$, but our best estimates at
0.2--1~kpc resolution imply very good agreement between this
low-metallicity dwarf and more massive disk galaxies.  Note that since
the scaling is linear, this result is insensitive to uncertainty in
inclination or other aspects of the SMC's geometry.  By contrast, the
total gas star formation law is offset significantly from that
observed in large galaxies. The SMC harbors unusually high \Shi\ and
low \Ssfr\ at a fixed total gas surface density (although the \Ssfr\
vs. \Sgas\ distribution moves closer to the loci of large spirals if
the star and gas are in a disk inclined by $i>40^\circ$, or if the
galaxy is elongated along the line of site). At a given \Sgas, the
molecular gas fraction is also offset to values lower than those
observed in massive disk galaxies, by typically an order of magnitude.

Two natural corollaries emerge from these observations. First,
molecular clouds in the SMC are not extraordinarily efficient at
turning gas into stars; star formation proceeds in them at a pace
similar to that in GMCs belonging to normal disk galaxies. This
suggests that, down to at least the metallicity of the SMC ($Z\sim
Z_\odot/5$), the lower abundance of heavy elements does not have a
dramatic impact on the microphysics of the star formation process,
although it does appear to have an important effect at determining the
fraction of the ISM capable of forming stars.

This is not a foregone conclusion. For example, it is conceivable that
the low abundance of carbon and the consequent low dust-to-gas ratio
and low extinction would affect the ionization fraction in the
molecular gas. This may result in changes in the coupling with the
magnetic field, perhaps slowing the GMC collapse and resulting in
lower \SFE\ and longer $\tau_{\rm dep}^{\rm mol}$. Or, alternatively,
the low abundance of CO (an important gas coolant in dense molecular
cores) could make it difficult for cores to shed the energy of
gravitational contraction, slowing their collapse and again resulting
in longer time scales for consuming the molecular gas \citep[but
see][]{KRUMHOLZ11}. Our result implies that to first order metallicity
does not have a large impact on the rate at which star formation
proceeds locally in molecular gas in the SMC. Firming up this conclusion,
however, requires detailed studies of molecular cloud lifetimes
\citep[for example, see][]{FUKUI99}.

Second, these observations provide very strong evidence that star formation
activity is related directly to the amount molecular gas, with
\hi\ coupled to SFR only indirectly. This should be tempered by 
the consideration that, as pointed out in \S\ref{subsec:H2map}, our
dust-derived \htwo\ map may include a contribution from very cold,
strongly self-absorbed \hi\ (such as that sometimes associated with
molecular cloud envelopes) which we cannot easily disentangle from
molecular material in our analysis.  The strong relation between
molecular material and star formation explains some puzzling
observations in the context of
\hi\ dominated systems. \citet{WOLFE06} searched for low surface
brightness galaxies in the Hubble Ultra Deep Field, with the
expectation that, based on Damped Lyman Alpha (DLA) neutral gas column
density statistics and the star formation law, a measurable fraction
of the sky should be covered by low brightness objects if star
formation takes place in DLAs. Their study shows that star formation
activity in DLAs is suppressed by over an order of magnitude (factors
of $30-100$) with respect to the predictions based on the total gas
Schmidt law. \citeauthor{WOLFE06} indeed suggest that part of the
explanation may be a low molecular fraction in DLAs. More recently,
\citet{KRUMHOLZ09b} show that although the observed distributon of
column densities in QSO-DLA systems theoretically requires the
existence of a significant cold phase, they are inconsistent with the
expectation of large molecular fractions. This is simply a reflection
of the fact that, given their low metallicities, their densities are
not large enough to sustain significant molecular fractions. The
analysis of the SMC presented here shows that star formation activity
is directly proportional to the molecular content.

\section{Comparison to Star Formation Models}
\label{sec:comparison}

How do our results in the SMC compare with predictions from models? 
The last few years have seen a range of very important theoretical and
computational modeling effort concerning the star formation law in
galaxies \citep[e.g.,][to mention just a
few]{SCHAYE04,BLITZ06,SCHAYE08,ROBERTSON08,GNEDIN10a,GNEDIN10b}.  In
the following sections we will focus on two recent theoretical models
for the dependence of star formation and phase balance on local gas
content and other galactic properties, those of \citet{KRUMHOLZ09},
henceforth KMT09, and \citet{OSTRIKER10}, henceforth OML10. Since the
models assume simple geometries, the reader should keep in mind the
caveats raised in \S\ref{subsec:smcintro} about the complex
line-of-sight geometry and overall structure of the SMC throughout
this section.

The models of KMT09 (summarizing a series of papers) and OML10 adopt
different simplifications and focus on different aspects of gas phase
balance in the ISM, and are therefore complementary.  The KMT09 model
adopts the simplification that all neutral gas is gathered into
high-density, cold atomic-molecular complexes with surface density
$\Sigma_{\rm comp}$ a factor $c$ larger than the mean gas surface
density \Sgas\ averaged over large scales (radio observations
typically sample scales $\gtrsim0.3$~kpc for galaxies outside the
Local Group).  The parameter $c$ is not predicted by KMT09, but the
comparison with observations suggests $c\sim5$ on scales
$r\sim750$~pc, and by definition $c\sim1$ on scales approaching the
size of molecular cloud complexes where $\Sigma_{\rm comp}$ dominates
the surface density of gas (several tens of parsecs to $\sim100$~pc for the
Milky Way).  Warm, diffuse \hi\ gas is assumed to represent a
negligible fraction of the total gas content.  This assumption does
not in fact appear to be satisfied globally in the SMC, based on our
estimated \htwo\ mass of $2.2 \times 10^7$
\msun\ and an \hi\ mass of $4.2 \times 10^8$ \msun, most of which is
believed to be warm \citep{DICKEY00}.  The cold \hi\ + \htwo\
component, however, probably still dominates over warm \hi\ locally.
In the KMT09 model, the \htwo/\hi\ balance within complexes is
determined by shielding of dissociating radiation, and depends
primarily on $Z \Sigma_{\rm comp}$.  Star formation is assumed to take
place exclusively within gas which is \htwo\ rather than \hi, at the
typical rates and efficiencies observed locally, of a few percent per
free-fall timescale. Comparison between our observations and KMT09 
are better carried out at the full resolution of the data, at
which the clumping-factor $c$ should approach unity.

The OML10 model adopts the simplification that all neutral gas is
divided between a diffuse component (consisting of both warm and cold
atomic gas in mutual pressure equilibrium), and a
gravitationally-bound component (consisting of cold atomic and
molecular gas in unspecified proportion).  The amount of diffuse
atomic gas is set by the requirement that heating (primarily FUV)
balances cooling, with the mean density of the diffuse medium (and
hence the cooling rate) set by dynamical equilibrium in the total
gravitational potential (provided by stars, gas, and dark matter).
Star formation, which produces the FUV, is assumed to occur at a
constant rate within the gravitationally-bound component. Because only
the very densest, highest-column gas within a gravitationally-bound
cloud actually forms stars and the distribution of densities and
columns in highly-turbulent clouds depends on temperature but not
\htwo\ content, OML10 assumed that the star formation rate in the
gravitationally-bound component is independent of the large-scale
\htwo/\hi\ proportions within this component, with a depletion time of
2 Gyr. Comparison to observations should be carried out on scales
where there is a mix of phases in equilibrium within a resolution
element, which we take to be approximately correct at $r\gtrsim 200$~pc.

\subsection{Comparison to KMT09}
\label{subsec:kmt09}

The KMT09 model is very successful at reproducing the composite
star-forming properties of local samples of galaxies with only three
inputs: surface density of gas, metallicity, and a clumping factor
$c\equiv \Sigma_{\rm comp}/\Sgas$ that approaches unity at high
spatial resolution.  The first two inputs are directly observable. On
the other hand, the clumping factor is poorly constrained in
unresolved observations, and is unlikely to be constant if
star-forming cloud complexes are gravitationally-bound entities that
are isolated from their environments.  The clumping factor is
introduced to correct the surface densities observed on large spatial
scales to the ``true'' gas surface density on GMC scales, $\Sigma_{\rm
comp}$.  Fortunately for us, on the scale of our SMC observations the
clumping factor should be approximately unity. Furthermore, these data
make it possible to test the effects of resolution on measurement of
the star formation law in \hi\ -dominated galaxies.

\begin{figure}[t]
\plotone{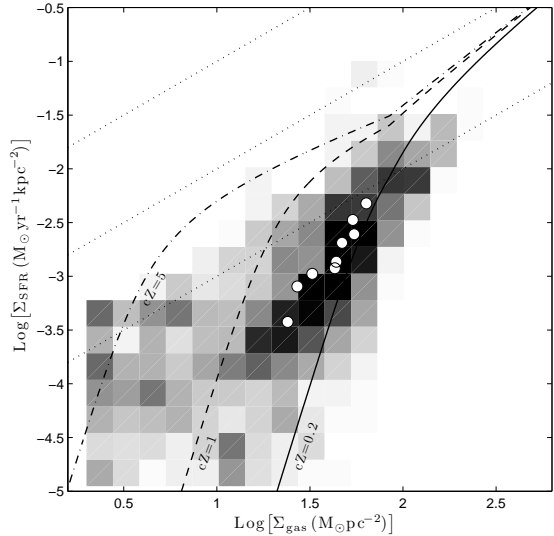}
\caption{Star formation law in the SMC at different spatial
resolutions (compare to the full-resolution results in Figure
\ref{fig:sflgas}). The gray scale shows the two-dimensional
distribution of the correlation between \Ssfr\ and \Sgas, as in Figure
\ref{fig:sflgas}, but now at a resolution $r\sim200$~pc. The white circles 
show the same information at $r\sim1$~kpc resolution. The results of
the KMT09 models for different values of the $cZ$ parameter are
indicated by the dash-dotted, dashed, and solid lines, as in Figure
\ref{fig:sflgas}. 
The dotted lines indicating constant SFE are as in
Figures \ref{fig:sflmol} and \ref{fig:sflgas}. Degrading the resolution tends
to move the star formation--total gas relation along a line parallel to 
the original distribution in Fig. \protect\ref{fig:sflgas}, 
rather than shifting to the left following 
the predictions of KMT09 for different clumping factors. 
\label{fig:fig6}\label{fig:resolution}}
\end{figure}

Figure \ref{fig:sflgas} shows the predictions from KMT09 overlaid on
the SMC data. We show the model results for three values of the
clumping factor--metallicity product, $cZ=0.2$, 1, and 5. The former
value corresponds to the metallicity of the SMC and a clumping factor
of unity, which is to be expected given our spatial resolution of
$r\sim12$ pc. The latter two values bracket most of the observations
originally used to test the model \citep{KRUMHOLZ09c}.  Although
the model drops off more steeply than the data at low \Sgas\ and
overshoots the observations at high \Sgas, the overall agreement
between the SMC observations and the $cZ=0.2$ model curve is very
reasonable.

\begin{figure}[t]
\plotone{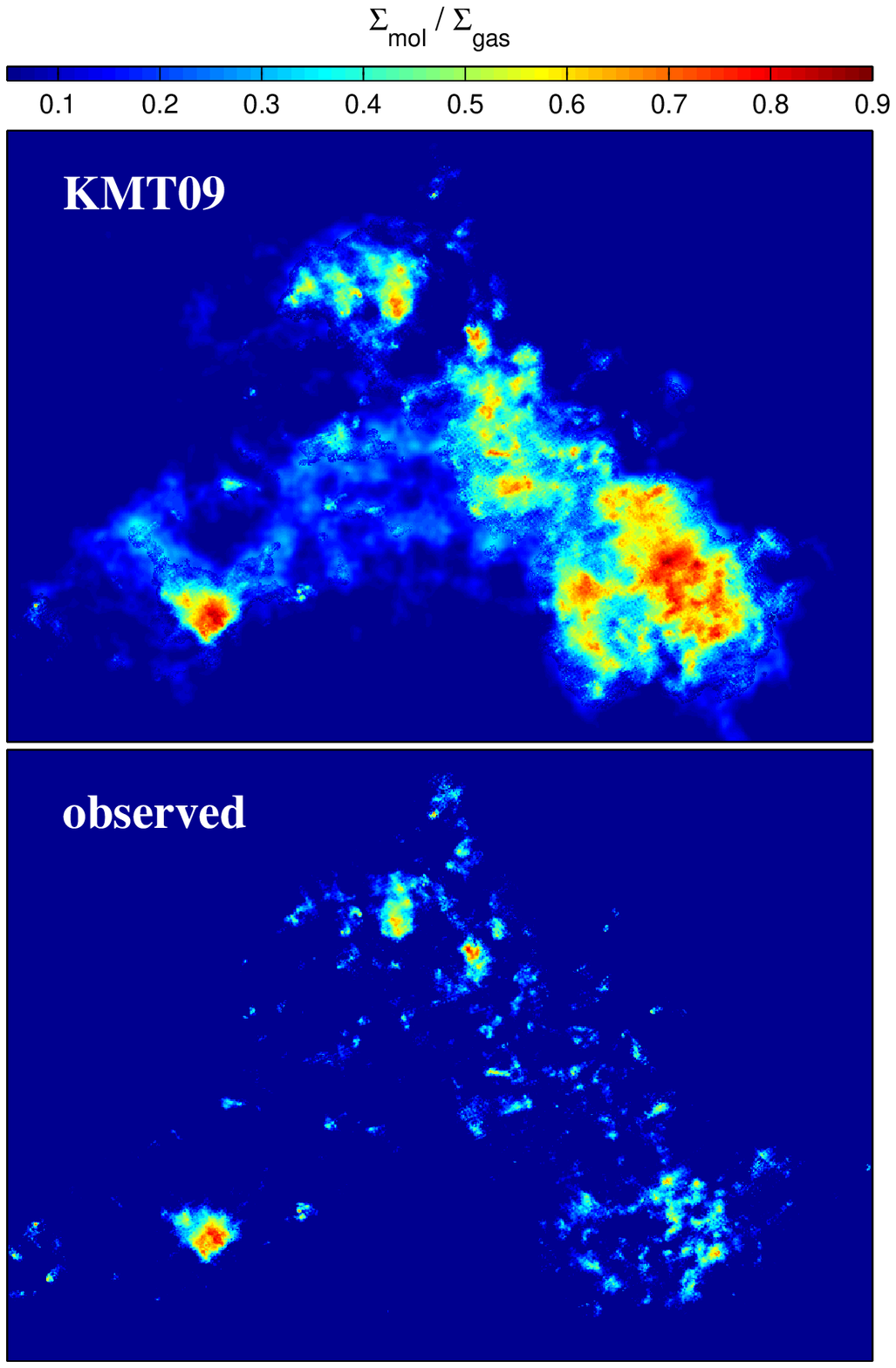}
\caption{Molecular fraction $\Smol/\Sgas$ in the SMC, 
comparison of KMT09 and measurements at 12~pc resolution.  The top
panel shows the results of KMT09 corresponding to our \Sgas\
measurements, computed for the metallicity of the SMC and unity
clumping factor. The bottom panel shows the measurements obtained from
the \htwo\ map presented here (Figure \ref{fig:h2map}) and the \hi\
observations by \citet{STANIMIROVIC99}. Although the agreement between
model and observations is reasonable, KMT09 tends to overpredict (by a
factor of $\sim2$) the molecular fraction (and the SFR) at high \Sgas.
This was already present in Figures \protect\ref{fig:fig3} and
\protect\ref{fig:fig5}, but it is much more apparent on a linear 
scale such as that used here. Note that, because of the spatial
filtering properties of the algorithm used to produce the \htwo\ map,
our measurements are not sensitive to an extended low level molecular
component.
\label{fig:fig7}\label{fig:kmtcomp}}
\end{figure}

KMT09 matches the star formation law observations reasonably well in
most part because of its success at reproducing the observed molecular
fraction as a function of column density. Figure \ref{fig:rH2} shows
the distribution of the ratio of molecular to atomic gas as a function
of total gas surface density at scales of $r\sim200$~pc, together with
the model tracks. Although most lines-of-sight are atomic-dominated,
the model with $cZ=0.2$ does a reasonable job at describing the
observations. There is almost no differentiation in the location of
different subregions of the SMC in this plot. Most notably, the SW end
of the SMC Bar with the largest surface densities (where most star
formation takes place) tends to lie preferentially below and to the
right of the model, while the NE end of the bar produces a lot of the
scatter toward lower column densities. The fact that KMT09 obtains
reasonable SFRs as a function of surface density suggests that one of
the fundamental assumptions underlying the model, that once gas turns
molecular it forms stars with approximately fixed efficiency and rate,
is approximately correct.

One important check on KMT09 is whether the clumping factor parameter
will explain the SMC data at different resolutions. Most of the
molecule-dominated disks used in previous comparisons with the model
have metallicities that are approximately solar and are consistent
with tracks having $cZ\sim1-5$, suggesting clumping factor values
$c\sim1-5$ at resolutions of several hundred parsecs. The results of
spatially smoothing the SMC data for \Ssfr\ and \Sgas\ to resolutions
$r\sim200$~pc and $r\sim1$~kpc are shown in Figure
\ref{fig:resolution}. The gray scale and the white circles represent
the results at 200~pc and 1~kpc respectively.  We can see that instead
of moving from the $cZ=0.2$ track at $r\sim12$~pc to the $cZ\sim1$
track at $r\sim1$~kpc the points tend to preserve the distribution of
the original data, sliding down along a line with slope $1+p\approx2$
(a fit to the 200~pc resolution points finds $1+p=2.1\pm0.2$). Thus
even at $r\sim1$~kpc the total gas star formation law points are found
more-or-less along the $cZ=0.2$ track.

Why is the inferred clumping-factor parameter not changing much with
resolution?  The main reason is that the distribution of \hi\ surface
density is very uniform in the SMC. Thus going from $r\sim12$~pc to
$r\sim1$~kpc does not dramatically affect the surface density in its
central regions. By contrast, the squares and circles in Figure
\ref{fig:fig2} show the effect of going to larger spatial scales on
the \htwo\ distribution.  The median of the logarithm of \Shtwo\ on
1~kpc scales decreases by $0.5-0.6$ dex from its value at
$r\sim12$~pc, about the $x$-axis separation between the $cZ=0.2$ and the
$cZ=1$ model tracks in Figure \ref{fig:resolution}. In other words, if
the distribution of \hi\ were as clumpy as the \htwo, the change in
spatial resolution would shift the points by about the separation
between the model tracks along the $x$-axis.  Thus, the \hi\ and \htwo\
gas does not follow a similarly clumped spatial distribution.  This
calls into question the assumption by KMT09 that the atomic medium is
primarily found in shielding envelopes of implicitly cold, dense \hi\
gas surrounding the \htwo\ gas.  Indeed, the observations of
\citet{DICKEY00} suggested that the \hi\ in the SMC is primarily warm,
diffuse gas rather than cold, dense clouds.

Note that $c\equiv\Sigma_{\rm comp}/\Sgas$, so using a constant $c$
over a range of surface densities is strictly incorrect, and this may
be partially the cause of the apparent slant of the data with respect
to the $cZ=0.2$ track, which is present at all resolutions. Impossing
$c\sim(\Sgas)^{-1}$, however, dramatically overcorrects this
effect. Fitting the observations requires a softer correction,
$c\sim(\Sgas)^{-0.5}$, suggesting a connection between the surface
density of cloud complexes and the density of the surrounding gas on
large scales. It is not immediately clear why this should be the case
in the context of KMT09.

It is worthwhile to note that although the KMT09 $cZ=0.2$ track is in
reasonable overall agreement with the data in Figure \ref{fig:fig5},
there is a noticeable bias toward overpredicting molecular ratios by
factors $\sim2-3$. Since the discrepancy is of the order of our
claimed systematic uncertainty for the \htwo\ map, this is most
meaningful in the sense of the relative comparison to OML10. Figure
\ref{fig:kmtcomp} shows the molecular gas fraction \Smol/\Sgas\
resulting from applying KMT09 to the observed \Sgas\ distribution at
full resolution (where we know the clumping-factor should be $c\sim1$
and $cZ\sim0.2$), compared to our measurements of the same
quantity. The discrepancy between model and observations in the
molecular fraction is most apparent for the SW tip of the Bar, which
harbors the largest atomic and total gas surface densities. Note that
the discrepancy between the data and the $cZ=0.2$ predictions persists
although somewhat diminished at $r\sim200$~pc (Figure \ref{fig:fig5}),
but the model applicable on those scales should have $c>1$ ($cZ>0.2$)
since the molecular complexes are likely unresolved and $\Sgas$ is
smaller than $\Sigma_{\rm comp}$.

\subsection{Comparison to OML10}

The model of OML10 was motivated in part by observations indicating
that the star formation rate and phase balance depend not just on the
gas surface density, but also on the density of the stellar component,
with the molecular content increasing roughly linearly with the
estimated pressure in the ISM \citep{BLITZ06,LEROY08}.  OML10
explained this empirical result in terms of the simultaneous need to
satisfy thermal and dynamical equilibrium in the volume-filling
diffuse component.  The cooling rate in the diffuse medium is
proportional to the gas density, which is controlled by vertical
dynamical equilibrium (and hence can be affected by the stellar
density if it is large).  The heating rate, associated with radiation
from high-mass stars in galaxies with active star formation, is
proportional to the amount of gas forming stars in gravitationally
bound cloud complexes (GBCs).  In order to balance cooling and
heating, an appropriate partition of gas into diffuse and
self-gravitating components is needed.  The solutions of the
simultaneous thermal/dynamical equilibrium equations were shown by
OML10 to agree very well with the radial profiles of star formation in
a sample of spiral galaxies.  The model of OML10 does not, however,
make a prediction for the atomic/molecular balance in the ISM; in \S
\ref{subsec:hidiffuse}, we describe an extension of the model that
provides this estimate, for comparison to the observations in the SMC.

We use the following implementation of OML10, based on their 
Equations 10 and $15-17$,

\begin{eqnarray}
\Sigma_{\rm diff} &=& x\Sigma_{\rm gas}, \label{eq:1}\\
\Sigma_{\rm gbc} &=& (1-x)\Sigma_{\rm gas}, \label{eq:2}\\
\Ssfr &=& \frac{(1-x)\Sigma_{\rm gas}}{\tau_{\rm dep}}, \label{eq:3}\\
\Pth &=& \frac{4\Ssfr}{\Sigma_{\rm SFR,0}}\times\frac{Z_d/Z_g}{1+3.1 (\Sigma_{\rm gas} Z_d/\Sigma_0)^{0.365}}, \label{eq:4}\\
\Sigma_{\rm diff} &=& \frac{9.5\msunperpcsq\,\alpha\Pth}{0.11\Sigma_{\rm gbc}+\left[{0.011(\Sigma_{\rm gbc})^2+\alpha\Pth+100\alpha\fw\rho_{\rm sd}}\right]^{1/2}}, \label{eq:sd}
\end{eqnarray}

\noindent where $x$ represents the fraction of gas in the diffuse phase, 
$\Sigma_{\rm gas}$, $\Sigma_{\rm diff}$, and $\Sigma_{\rm gbc}$ are
the large-scale averages of the surface densities of total gas,
diffuse phase, and gravitationally-bound cold phase respectively (all
in \msunperpcsq), and $\rho_{\rm sd}$ is the midplane volume density of
stars plus dark matter (in \msunperpccu). The parameters
$\Sigma_0\approx10\,\msunperpcsq$, $\Sigma_{\rm
SFR,0}\approx2.5\times10^{-9}\,\msunperpcsqyr$, $\alpha\approx5$,
$\fw\approx0.5$, and $\tau_{\rm dep}\approx2\times10^9$~yr are
respectively the surface density of gas and the star formation rate at
the Solar circle, the ratio of total to thermal pressure, the fraction
of diffuse gas in the warm phase, and the gas depletion time in the
gravitationally-bound component (as inferred empirically from spiral
galaxies).  The parameter $\Pth$ is simply the normalized thermal
pressure in the model, $\Pth\equiv(P_{\rm th}/k)/3000\,\mbox{\rm K
cm$^{-3}$}$ where $k$ is Boltzmann's constant.
For a two-phase ISM, $P_{\rm th}$ is proportional to the UV heating rate 
\citep{WOLFIRE03}, which 
is proportional to \Ssfr\ (as expressed in Equation \ref{eq:4}).
Finally, the dust-to-gas ratio and the gas phase metallicity, both
relative to the Milky Way Solar circle values, are indicated by $Z_d$
and $Z_g$. For the SMC we assume $Z_d\cong Z_g=0.2$
\citep[e.g.,][]{LEROY07}.

Given a local $\Sigma_{\rm gas}$ and $\rho_{\rm sd}$, we substitute Equations
\ref{eq:1} to \ref{eq:4} into Equation
\ref{eq:sd}, finding the value of $x$ in the $[0,1]$ interval that
satisfies it. The corresponding value of \Ssfr\ can then be readily
found using Equation \ref{eq:3}. 

As inputs to the model, besides our total surface density gas map, we
employ the stellar surface density derived from the SAGE-SMC 3.6
$\mu$m images \citep{GORDON11} using the mass-to-light ratio from
\citet{LEROY08}, and the dark matter density profile results from
\citet{BEKKI09}. To deproject the stellar surface density we use a
disk scale height of 2~kpc. Because the thickness of the SMC is very
poorly constrained this is simply a guess. Nonetheless, since the term
associated with $\rho_{\rm sd}$ in Equation \ref{eq:sd} is much
smaller than that associated with the gas terms, it turns out that the
precise value of the scale height has little impact on the final
result.  Under these assumptions the typical values for the bar and
wing regions of the SMC are $\rho_{\rm
sd}\simeq0.015-0.03$~\msunperpccu, with a dark matter contribution
$\rho_{\rm dm}\simeq0.008$~\msunperpccu. As explained in OML10, for
near-Solar metallicity and GBCs with cloud complexes surface densities
$\Sigma_{\rm comp} \sim 100$ \msunperpcsq, individual GBCs are
well-shielded so that $\Shtwo\sim\Sgbc$ and $\Shi\sim\Sdiff$.  At low
metallicity, however, GBCs may have substantial \hi\ envelopes such
that $\Shtwo<\Sgbc$ and $\Shi>\Sdiff$. We will revisit this issue in
\S\ref{subsec:hidiffuse}, obtaining an estimate for the separate \hi\
and \htwo\ contributions to \Sgbc.

We find that following equations \ref{eq:1} to \ref{eq:sd} with
the same parameters as adopted by OML10, \Sdiff\ is substantially
lower than \Shi\ in the SMC, yielding \Sgbc/\Sdiff\ much larger than
the observed \Shtwo/\Shi\ ratios.  The distribution of the model
predictions computed for $\rho_{\rm sd}$ in the SMC are shown by the gray
contour in Figure \ref{fig:rH2}. This result is robust to our choice
of $\rho_{\rm sd}$ and the precise values of $\fw$, $Z_d$, and
$\tau_{\rm dep}$. In essence, total gas pressure (which is $\alpha\,P_{\rm
th}$) is what balances self-gravity, and $P_{\rm th}$ must also be
consistent with the balance between heating and cooling.  The
equilibrium value of $P_{\rm th}$ is insensitive to metallicity
(Equation \ref{eq:4}) because photoelectric heating is $\propto Z_d$
whereas metal cooling is $\propto Z_g$.  Consequently at the high
column densities and pressures present in the SMC, the model requires
that most of the gas is driven into the self-gravitating cold phase
almost independent of metallicity, in order to generate the radiation
field and the corresponding heating needed to attain pressure
equilibrium in the diffuse gas. For \Sdiff\ to approach the large
values of \Shi\ observed in the SMC would imply large values of the
pressure, which necessitates either increases in $P_{\rm th}$ (and
consequently the heating), or a large turbulent factor represented by
$\alpha$. Matching the observations with an increase in $\alpha$
requires a very large value, $\alpha\sim20$, which is likely
unrealistic \citep{BURKHART10}. For the reference value,
$\alpha\approx5$, the predicted surface density of the diffuse phase
in the SMC is \Sdiff$\lesssim30$ \msunperpcsq, in contrast with the
higher range of \Shi\ evident in Figure
\ref{fig:sflgas}.

It is important to realize that this point is pretty much independent
of the details of the model. In any scheme where the gas is in
pressure equilibrium and the geometry is not pathological, sustaining
large diffuse-\hi\ surface densities requires a correspondingly large
heating term to counterbalance the strong cooling of warm \hi\ at high
pressure and density.  The gas heating can be driven by the radiation
field, or else dynamical energy input (as in the example of increasing
the $\alpha$ parameter) can puff up the disk to reduce its density and
cooling rate.  In either case, energy and momentum input from star
formation needs to be present to provide the support necessary to
prevent collapse. To match the typical \hi\ surface density observed
in the SMC at 200~pc resolution,
\Shi$\sim60-70$ \msunperpcsq,  would require
a thermal pressure
$P_{\rm th}/k\sim24-32\times10^4$ K~cm$^{-3}$ in the diffuse phase (assuming
$\alpha=5$), far in excess of what is observed in the Milky Way plane.

Attaining such pressures requires a large heating term, and
equilibrium demands it be balanced by an equally large cooling. Since
the cooling in neutral gas is dominated by far infrared fine structure
lines over a wide range of densities \citep{WOLFIRE95}, we expect that
the consequence of increased heating in the diffuse medium would be
observable. Note that there is evidence that the \cii\ 157.7 $\mu$m
transition is indeed bright in the Magellanic Clouds and other low
metallicity dwarfs \citep{STACEY91,ISRAEL96,MADDEN97}, but those
measurements are likely dominated by the dense, self-gravitating
phase. Testing this hypothesis will likely require observations
of the diffuse phase.

Before proceeding to consider the possibility of enhanced heating in
more detail, we note that an increase in the assumed inclination angle
$i$ of the SMC's plane would also yield a lower prediction for
\Sgbc/\Sdiff, and lower $P_{\rm th}$.  For the gas-gravity-dominated case,
Equation (\ref{eq:sd}) yields $y \sim (\Sigma_{\rm gas}/9.5
M_\odot\ {\rm pc}^{-2})^2/\alpha$ for the normalized thermal pressure,
while Equations (\ref{eq:2})-(\ref{eq:4}) yield $y\sim \Sigma_{\rm
  gbc}/(\Sigma_{\rm SFR,0} \tau_{\rm dep})$.  If diffuse gas
dominates, combining these leads to
\begin{equation}
\frac{\Sigma_{\rm gbc}}{\Sigma_{\rm diff}}\sim
\frac{\Sigma_{\rm gbc}}{\Sigma_{\rm gas}}\sim 
\frac{\Sigma_{\rm gas} \Sigma_{\rm SFR,0} \tau_{\rm dep}}
{(9.5 M_\odot\ {\rm pc}^{-2})^2 \alpha}.
\end{equation}
Taking fiducial parameter values for $\Sigma_{\rm SFR,0}$, $\tau_{\rm dep}$, and 
$\alpha$, this yields
$\Sigma_{\rm gbc}/\Sigma_{\rm gas}\sim 
\Sigma_{\rm gas}/100 M_\odot\ {\rm pc}^{-2}$.  A reduction of deprojected 
\Sgas\ by a factor $\mu$ would lower \Sgbc/\Sdiff\ by the same factor.
Thus, if the typical deprojected value of \Sgas\ were 20 \msunperpcsq\
rather than 60 \msunperpcsq, the predicted \Sgbc/\Sdiff\ would be a
factor $\sim 3$ lower.  To reduce \Sgbc/\Sdiff\ to the level $\sim
0.1$ of the molecular-to-atomic ratio would require an inclination
correction factor $\mu \sim 0.1$ (corresponding to an almost edge-on
inclination $i\sim85^\circ$), however, which seems unlikely.

\subsubsection{Enhanced heating?}
\label{subsec:heating}

What could be the source of the extra diffuse gas heating in the case
of the SMC? In the following we explore the possibility of more
effective diffuse gas heating happening at low metallicities.  One
potential solution is that the local radiation field (relative to the
Solar neighborhood), $G_0'$, is enhanced at low metallicities with
respect to the corresponding case at $Z=1$. A simplification
introduced in the OML10 model, as described in the paper, is to ignore
local variations in the propagation of the FUV photons, which
determine $G_0'$ and the heating of the gas. Thus, the radiation field
relative to the Solar circle in the Milky Way is represented in the
heating term (Equation
\ref{eq:4}) by the factor $\Ssfr/\Sigma_{\rm SFR,0}$. Note that the
optical depth to FUV radiation, $\tau_{\rm UV}$, is determined by dust
attenuation. For identical columns of gas, this is lower in a low
metallicity environment because of the lower dust-to-gas ratio $Z_d$,
such that $\tau_{\rm UV}\approx 1/2\,A_\lambda/A_V\,(N_{\rm
H}/2\times10^{21}
\percmsq )\,Z_d$. The escape probability of FUV photons, $\beta_{\rm
UV}\simeq(1-e^{-\tau_{\rm UV}})/\tau_{\rm UV}$, which in the limit of
large $\tau_{\rm UV}$ is $\beta_{\rm UV}\sim1/\tau_{\rm UV}$, is
consequently higher at lower $Z_d$ (for a given $N_H$).  To account
for this effect we can introduce a factor of $\beta_{\rm UV}\sim1/Z_d$
in Equation \ref{eq:4}, so that the normalized thermal pressure
becomes

\begin{equation}
\Pth = \frac{4\Ssfr}{\Sigma_{\rm SFR,0}}\times\frac{1/Z_g}{1+3.1 (\Sigma_{\rm gas} Z_d/\Sigma_0)^{0.365}}. \label{eq:pth}
\end{equation}

Interestingly, there is some evidence for an enhanced mean radiation
field in the Bar region of the SMC. Modeling of the dust emission
suggests that the average $G_0'$ is $\sim4-5$ times larger than the
local \citet{MATHIS83} radiation field
\citep{SANDSTROM10}. An extended warm dust component associated with the 
Bar is also observed in the recent analysis using {\em Planck} data by
\citet{PLANCKCOLLABORATION11}.  The mean radiation field in the Wing by 
\citet{SANDSTROM10} appears to be approximately Galactic, with the 
exception its tip and the star forming regions there located, despite
the large values of \Shi\ there present. The modeling of the dust
emission, however, is particularly difficult in the Wing, where the
signal-to-noise of the observations is lowest.

\begin{figure}[t]
\plotone{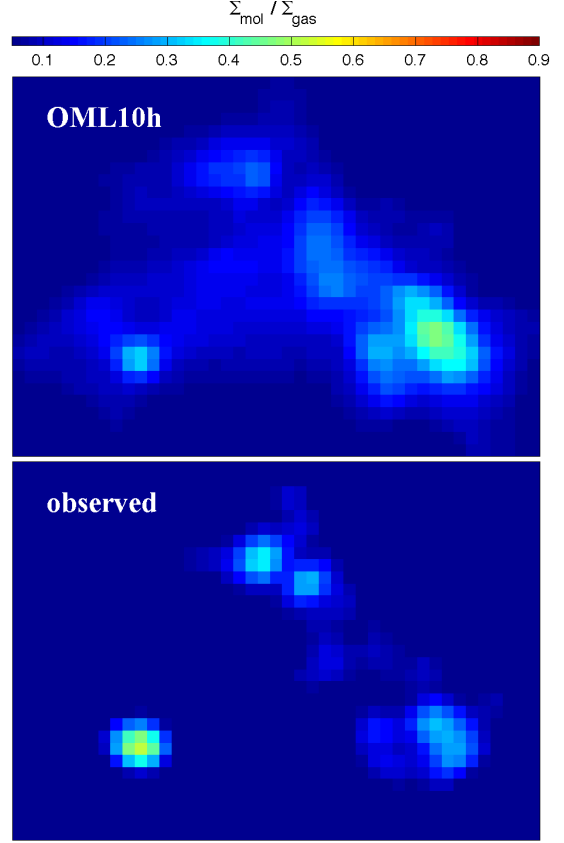}
\caption{Molecular fraction $\Smol/\Sgas$ in the SMC, 
comparison of OML10 and measurements at 200~pc resolution.  The top
panel shows the results of OML10h, the OML10 model modified to include
enhanced heating in the diffuse phase at low metallicities as discussed in
\S\ref{subsec:heating}. The bottom panel shows the measurements
obtained from the \htwo\ map presented here (Figure \ref{fig:h2map})
and the \hi\ observations by \citet{STANIMIROVIC99}. Note that,
because of the spatial filtering properties of the algorithm used to
produce the \htwo\ map, our measurements are not sensitive to an
extended low level molecular component.
\label{fig:fig8}\label{fig:omlcomp}}
\end{figure}

With Equation \ref{eq:pth} replacing Equation \ref{eq:4} in the model
the resulting \Sdiff\ reproduces the large observed \Shi\ in the SMC
using the nominal (Galactic) values for the rest of the parameters,
including $\alpha$. A comparison of the predictions of OML10 modified
for heating (OML10h henceforth) with our measurements of the molecular
fraction are shown in Figures \ref{fig:fig5} and \ref{fig:omlcomp}.
OML10h does very well at reproducing the average trends in the data,
although local over- or under-predictions persist at the factor of two
level.

The above approach is perhaps unrealistically simple, and it is likely
to overestimate the correction to the local interstellar radiation
field due to the diminished FUV extinction at low metallicities.  An
important fraction of the UV radiation escapes HII regions through
very low extinction lines-of-sight, and that fact should be taken into
account in the FUV propagation. In more general terms, the local
interstellar radiation field relative to the Solar neighborhood,
$G_0'$, is
\begin{equation}
G_0' = \frac{(1-x)\Sigma_{gas}}{\tau_{\rm dep}\Sigma_{\rm SFR,0}}\frac{f}{f_0},
\end{equation}

\noindent where $f/f_0$ is the enhancement factor for the radiation 
field relative to the Solar circle value due to propagation effects, and
it includes the escape of radiation from star-forming regions, as well
as the propagation through the diffuse gas. This factor can be written
as

\begin{eqnarray}
\frac{f}{f_0} &=& \left[\frac{f_{\rm esc}+\frac{1-e^{-\tau_{\rm gbc}}}{\tau_{\rm gbc}}
(1-f_{\rm esc})}{f_{\rm esc,0}+\frac{1-e^{-\tau_{\rm gbc,0}}}{\tau_{\rm gbc,0}}(1-f_{\rm esc,0})}
\right] \times 
\nonumber \\
& & \hskip 2cm
\left[\frac{1-E_2(\tau_{\rm diff}/2)}{1-E_2(\tau_{\rm diff,0}/2)}\frac{\tau_{\rm diff,0}}{\tau_{\rm diff}}\right],
\label{eq:fcor}
\end{eqnarray}

\noindent where $f_{\rm esc}$ is the fraction of the UV radiation 
directly escaping star-forming regions, $E_2$ is the second
exponential integral function, $\tau_{\rm gbc}$ and $\tau_{\rm diff}$ are the
optical depth to the UV associated with gravitationally bound clouds
and diffuse gas, respectively, and the 0 subscript indicates Solar
circle Milky Way reference values.  The second term in square
brackets, as introduced in OML10, asssumes the diffuse gas is a
uniform slab with total optical depth $\tau_{\rm diff}=\kappa_{\rm UV}
\Sigma_{\rm diff}$. Note that the correction factor introduced in
Equation \ref{eq:pth} is just a simplification of Equation
(\ref{eq:fcor}) considering only the factor between the first square
brackets, due to the gravitationally bound phase, with
$f_{\rm esc}=0$. Although it is possible to reproduce the observations of
the SMC using this approach, it requires choosing values for a number
of very poorly constrained parameters.

Other poorly constrained parameters in the problem of how the low
metallicity affects the diffuse gas heating are the changes in the
properties of the dust grains that couple the radiation field to the
gas.  The smallest carbonaceous dust grains in the ISM, associated
with the 2175 \AA\ extinction bump and the mid-infrared aromatic
features attributed to polycyclic aromatic hydrocarbons
\citep[PAHs;][]{DRAINE01,LI02}, are considerably depleted at low
metallicities and in particular in the SMC
\citep{MADDEN06,ENGELBRACHT08,SANDSTROM10}. If these grains dominate
the heating, as it appears to be the case for the LMC
\citep{RUBIN09}, the consequently lower photoelectric heating
efficiency could make it more difficult to heat the diffuse gas despite
the enhanced radiation field.

We conclude that including systematic heating effects in the diffuse
\hi\ gas at low metallicities, due to the enhanced radiation escape
from the dense star-forming phase offers a viable model, able to
approximately reproduce the data. We will return to whether this
is the best solution in \S\ref{subsec:consequences}.

\begin{figure*}[t!]
\plottwo{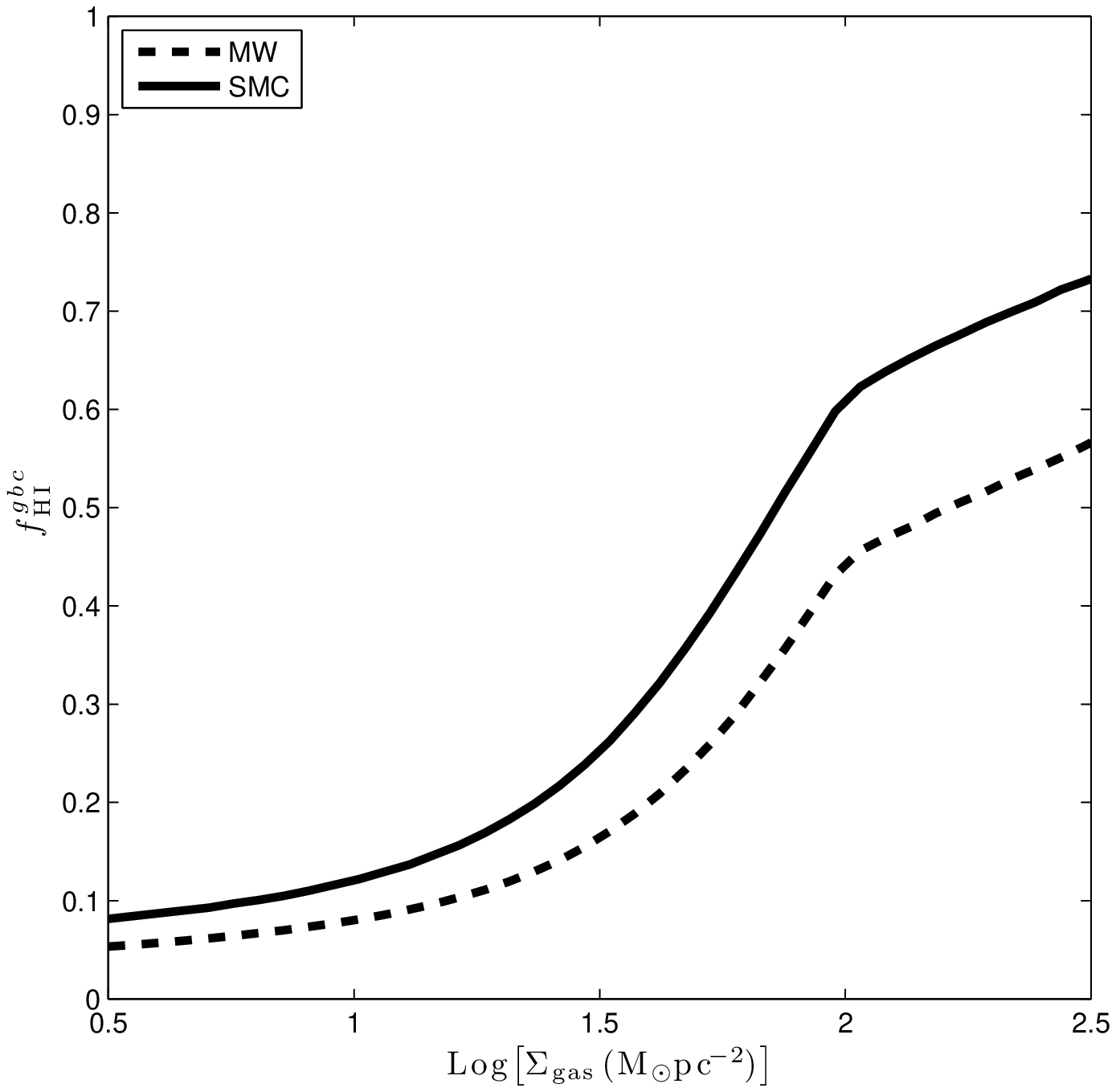}{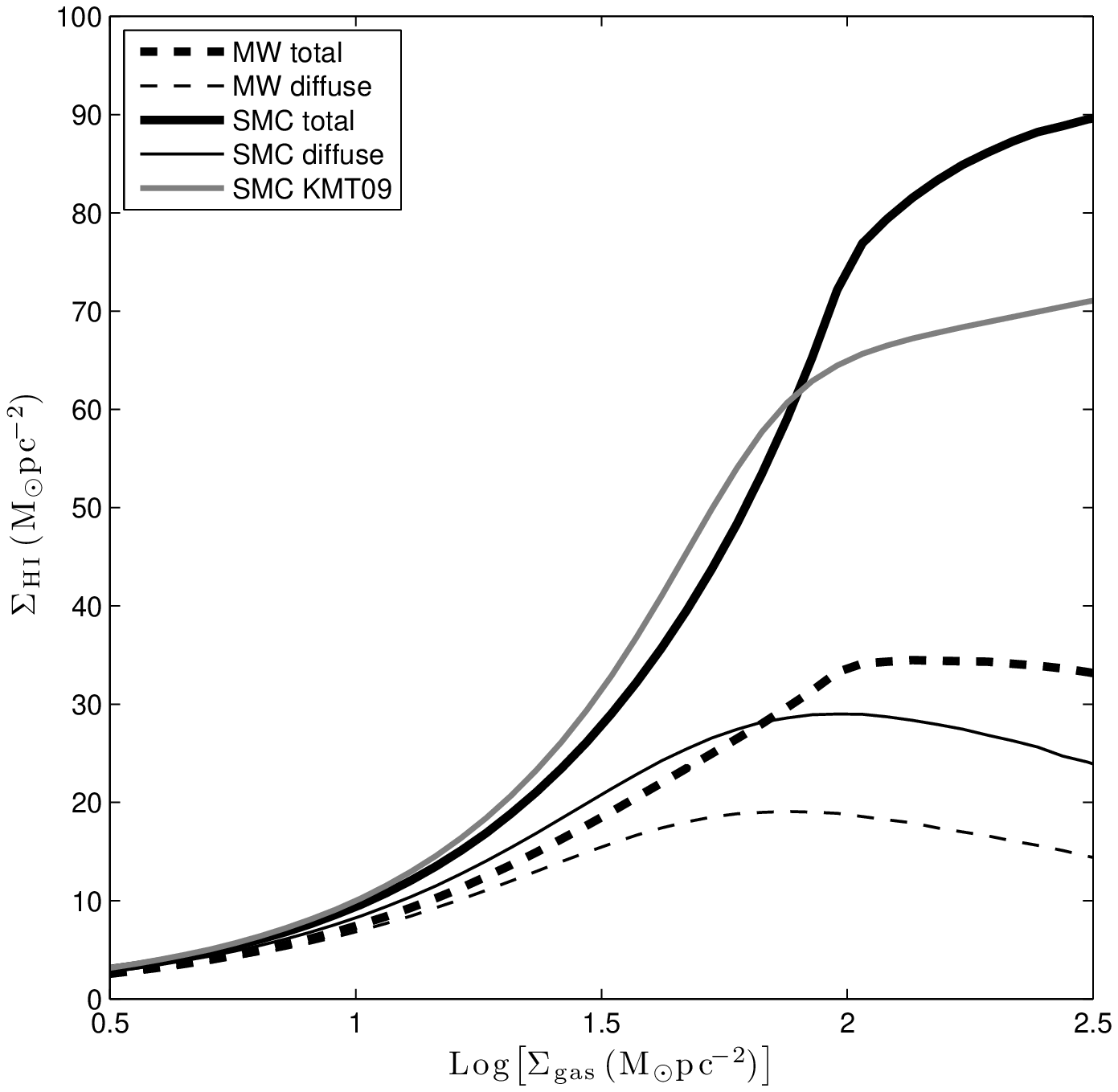}
\caption{Behavior of the atomic component in the OML10p model described in 
\S\ref{subsec:hidiffuse} and comparison to the diffuse component in OML10, 
under typical conditions for 
the Solar circle in
the Milky Way ($\rho_{\rm sd}=0.05$
\msunperpccu\ and $Z=1$) and the SMC ($\rho_{\rm sd}=0.02$ \msunperpccu\
and $Z=0.2$).  {\em (Left)} Fraction of the total \hi\ surface density
arising from the gravitationally bound phase according to
Eq. \ref{eq:fgbc}. The diffuse and the
gravitationally bound phases make approximately equal contributions to
\Shi\ at surface densities of approximately 160 (70) \msunperpcsq\ for typical 
conditions in the local Milky Way (SMC). {\em (Right)} Surface density
of \hi\ as a function of total gas surface density in OML10p, with the
prediction of KMT09 for comparison. With the contribution of \hi\ from
the gravitationally bound phase, it is possible to match the
typical $\Shi\approx70$ \msunperpcsq\ observed in the SMC without an
increase in the heating term.
\label{fig:fig9}\label{fig:hiunified}}
\end{figure*}

\subsubsection{Is all the HI diffuse?}
\label{subsec:hidiffuse}

An alternative explanation for the observations of the SMC in the
context of the OML10 model is that only a small fraction of the
observed \Shi\ is diffuse gas contributing to \Sdiff, and the
remainder of the \hi\ is actually in gravitationally bound structures
contributing to \Sgbc.  Physically, we would expect the \hi\ component
of individual GBCs to be the envelopes, and the \htwo\ component the
shielded interiors (note that the clumpy structure of clouds means
both the ``envelope'' and ``interior'' may pervade the whole cloud
complex). The relation between \hi\ and \htwo\ within GBCs would then
be determined by the physics of photodissociation described in the
KMT09 model.  In principle it should be possible to test this idea by
searching for the kinematical signatures of these gravitationally
bound structures in the existing \hi\ datacubes.

As noted in \S \ref{subsec:kmt09},
we have argued in \S\ref{subsec:gassfr} that neither Figure
\ref{fig:cumdist} nor the results of \citet{DICKEY00} support 
the scenario that most of the \hi\ is cold gas in the envelopes of
GBCs (which are atomic-molecular cloud complexes in the terminology of
KMT09).  Nonetheless, it is interesting to further investigate its
consequences in the context of the OML10 model, just as we did for the
model of KMT09.

To compute the amount of \hi\ in GBC envelopes we can apply the
molecular fraction determination from the KMT09 model to the
self-gravitating portion of the gas, and rely on the OML10 model
described by Eqs. \ref{eq:1}--\ref{eq:sd} to determine the balance
between the diffuse and self-gravitating phases.  The equations for
the resulting surface densities of \hi\ and \htwo\ averaged over large
scales are

\begin{eqnarray}
\Shi &=& \Sgas \left[x+(1-x)(1-f_{\rm H2})\right], \label{eq:shi} \\
\Shtwo &=& \Sgas (1-x)f_{\rm H2}, \label{eq:sh2}
\end{eqnarray}

\noindent where 
$f_{\rm H2}=f_{\rm H2}(\Sigma_{\rm comp},Z)$ is determined through
Equation 2 in KMT09. It is a function of the surface density of the
cloud, $\Sigma_{\rm comp}$, and the metallicity $Z$ relative to the
Milky Way, which we take to be $Z=Z_d=Z_g$. We take $\Sigma_{\rm
comp}$, to be the larger of $\Sgbc=(1-x) \Sgas$ and $100$
\msunperpcsq. The latter is the approximate value for the
surface density of resolved molecular clouds in nearby galaxies
\citep{BOLATTO08}. Thus this model assumes that 
GBCs have at least this surface density, and lower surface densities
are the result of beam dilution in the observations. The fraction of
all \hi\ found in the gravitationally bound phase, $f_{\rm HI}^{\rm
gbc}$, will then be

\begin{equation}
f_{\rm HI}^{\rm gbc} = \frac{(1-x)(1-f_{\rm H2))}}{\left[x+(1-x)(1-f_{\rm H2})\right]}. \label{eq:fgbc}
\end{equation}

Because this model extends OML10 by incorporating the
photodissociation model of KMT09 for comparison to observations of the
\hi\ abundance, we henceforth refer to it as OML10p.  The behavior of
the \hi\ component in this model is shown in Figure
\ref{fig:hiunified}. After accounting for the \hi\ in the
gravitationally bound gas through the inclusion of the
photodissociation term from KMT09, the surface density of atomic gas
saturates at $\Shi\sim90$ \msunperpcsq\ under the typical conditions
found in the SMC instead of the $\Shi\lesssim30$ \msunperpcsq\
stemming from the diffuse component alone. In fact, the \hi\
associated with the gravitationally bound gas becomes the dominant
contributor to
\Shi\ for surface densities $\Sgas\gtrsim70$
\msunperpcsq\ under SMC conditions. Inclusion of the 
modifications to the heating term discussed in \S\ref{subsec:heating}
mostly affect this threshold surface density, and increase the
saturation value of \Shi\ only moderately ($\Shi\lesssim150$
\msunperpcsq\ using Equation
\ref{eq:pth}, for example).

\begin{figure}
\plotone{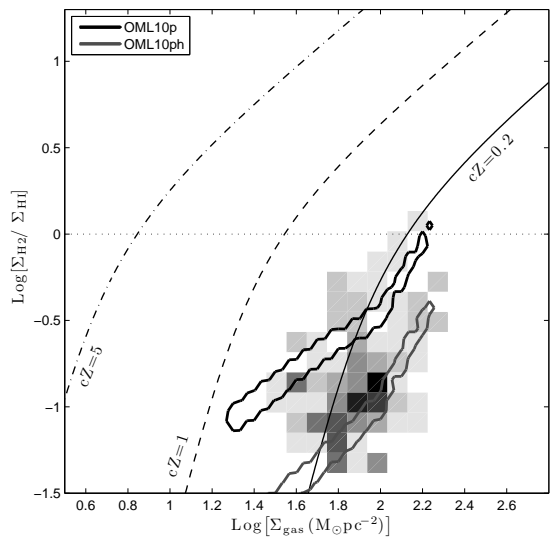}
\caption{Ratio of molecular to atomic surface densities in the SMC compared to
model calculations. The thick black contours show the predictions of
the OML10p model discussed in \S\ref{subsec:hidiffuse}, where we
include the photodissociation of \htwo. In this model, \Shi\ in the
SMC has a dominant contribution from \hi\ in the gravitationally bound
phase. The prediction of the OML10ph model, which includes both
photodissociation and enhanced heating in the diffuse phase (see \S
\ref{subsec:unified}), is also shown.  Lines from KMT09 are as in
Figure \ref{fig:rH2}. \label{fig:fig10}\label{fig:unirH2}}
\end{figure}

Figure \ref{fig:unirH2} shows the distribution of the molecular to
atomic ratio $\Shtwo/\Shi$ in the SMC according to this model,
compared with KMT09 and the distribution of the measurements at 200 pc
resolution as in Figure \ref{fig:rH2}. The similarity at \Sgas
$\gtrsim 100$\msunperpcsq\ between OML10p and the KMT09 track
corresponding to $cZ=0.2$ is simply due to the fact that the
\hi\ arising from the diffuse component is not dominant at 
these surface densities. The inclusion of the diffuse \hi\ in the
\Shi\ budget is responsible for the displacement down from the KMT09
track by a factor of $\sim0.6$.

We conclude that including the effects of photodissociation into the
calculations to compare the predictions of OML10 to the \hi\ and
\htwo\ data is physically motivated and potentially offers an
alternative explanation for the observed low values of the
$\Shtwo/\Shi$ ratio.  As we discuss in the next section, however, we
think this is not the dominant consideration in explaining the data.

\subsubsection{Consequences for star formation}
\label{subsec:consequences}

We have mentioned possible tests of these ideas looking for evidence
for enhanced heating or \hi\ in the GBC phase using: 1) further
modeling of dust observations to determine the radiation field, 2)
observations of diffuse ISM coolants such as \cii, and 3) searching
for kinematic signatures of gravitationally bound \hi. It should be
possible, however, to make an effort to distinguish between these
enhanced-heating and standard heating variation of OML10 model by
their impact on the star formation law.

In the context of OML10, all of the gas in the gravitationally bound
component could potentially contribute to star formation, although due
to the turbulent dynamics within GBCs, only a tiny fraction ($\sim
1\%$ per free-fall time) is dense (and cold) enough that it actually
collapses and forms stars.  In the variant of the model with enhanced
heating, OML10h, the gravitationally bound phase is depressed to allow
more diffuse \Shi. In the extension of the model that includes an
estimate for the contribution to \Shi\ from photodissociated envelopes
of GBCs, OML10p, the gravitationally-bound component is about as
abundant as at higher metallicities, and the low observed
$\Shtwo/\Shi$ ratio is explained by limited shielding of the
gravitationally bound gas at low metallicity (as for the dense
atomic-molecular complexes in KMT09). As a consequence of the higher
abundance of gravitationally-bound gas, the SFR expected in OML10p at
a fixed gas surface density is significantly higher than in OML10h.

\begin{figure}[h!]
\plotone{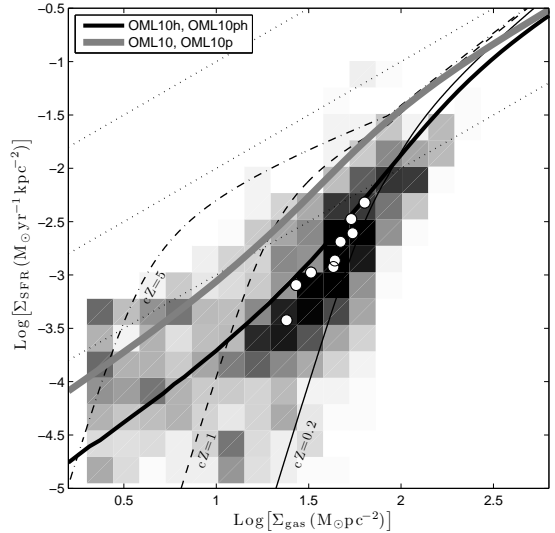}
\caption{
Comparison of OML10 predictions and total gas star formation law
observations in the SMC. The thick solid black and gray lines show the
results of the enhanced-heating (OML10h, OML10ph) and original-heating
(OML10, OML10p) versions of the model.  As in Figure
\ref{fig:resolution} the gray scale and white circles represent the
SMC observations smoothed to a resolution of 200 pc and 1 kpc
respectively. The KMT09 tracks and the constant gas depletion time
lines are also as in Figure
\ref{fig:resolution}. \label{fig:fig11}\label{fig:sfroml}}
\end{figure}

Figure \ref{fig:sfroml} shows the SFR predicted by OML10 with and
without additional heating (taking $\rho_{\rm sd}=0.02$\msunperpccu and
$Z=0.2$ for the SMC), compared to the observations smoothed to 200 pc
and 1 kpc resolution and to the KMT09 model, as in Figure
\ref{fig:resolution}. Because the specific star 
formation rate in GBCs is assumed to be insensitive to their exact
atomic/molecular balance, there is no difference in the predicted star
formation rate between OML10 and OML10p.  Similarly, OML10h and
OML10ph (see below) have the same
\Ssfr\ because they have the same \Sgbc\ for a given \Sgas.  The
predictions for \Ssfr\ from OML10, OML10p, and KMT09 with $c=5$ are
quite similar for \Sgas $\gtrsim 20$\msunperpcsq, because for all of
these models in this range of \Sgas, the majority of gas is
concentrated in cold, dense, star-forming atomic-molecular complexes
with $\Sigma_{\rm comp} \gtrsim 100$\msunperpcsq.  These models
overpredict the \Ssfr\ observations by about a factor of 6. On the
other hand, OML10h (and OML10ph) do a very good job at matching not
only the average level of SFR activity but also the slope of the
\Ssfr--\Sgas\ relation, despite the simplicity of the heating
correction.  

\subsubsection{A unified model}
\label{subsec:unified}

It is possible to think of further modifications to OML10. For
example, we can extend the enhanced-heating model to include an
estimate of the atomic/molecular balance within GBCs; we refer to this
as OML10ph. As noted above, this will produce the same
\Ssfr-\Sgas\ results as the OML10h model as it has the same fraction
of gravitationally bound gas, but it will yield a molecular to atomic ratio
\Shtwo/\Shi\ a factor $\sim2.8$ times lower than OML10p (equivalent
to a displacement of $\sim0.45$ in the logarithm; as shown in Figure
\ref{fig:unirH2}). Such a model is not forbidden by the observations,
and in fact it arguably reproduces better the lower values in the
distribution of molecular to atomic ratios than the model without
enhanced heating.  In model OML10ph, it is also possible to compute
the fraction of \hi\ gas associated with the gravitationally bound
component, $f^{gbc}_{\rm HI}$, which has a value $f^{gbc}_{\rm HI}\sim
19\%$ at a surface density $\Sgas=100$
\msunperpcsq, very similar to the 15\% cold \hi\ fraction measured by
\citet{DICKEY00}.

\section{Summary and Conclusions}
\label{sec:conclusions}

We present a detailed analysis of the correlation between the atomic
and molecular gas phases in the SMC, and the star formation activity.
This is the first time such analysis has been done at high spatial
resolution in a galaxy of such a low metallicity. We use modeling of
the {\em Spitzer} dust continuum observations
\citep{BOLATTO07,GORDON11} to calculate the \htwo\ surface densities
on scales of $\theta\sim40\arcsec$ or $r\sim12$~pc, together with
observations of the \hi\ \citep{STANIMIROVIC99} and \ha\
\citep{SMITH99} distributions (Figure \ref{fig:h2map}). An important
caveat to keep in mind when considering the interpretations of the
results is the complexity of the \hi\ distribution in the SMC,
referred to in \S\ref{subsec:smcintro}. Another important caveat are
the uncertainties involved in the production of the \htwo\ map, which
we discuss in \S\ref{subsec:H2map}.

We find that, in the regions where we detect \Shtwo, we measure a
typical molecular gas depletion time $\tau_{\rm dep}^{\rm mol}\sim7.5$
Gyr on scales $r\sim12$~pc with a factor of 3.5 uncertainty accounting
for the scatter as well as the systematics associated with our \htwo\
map as well as the geometry of the source (Figure
\ref{fig:sflmol}). The depletion time shortens when measured on larger
spatial scales. On scales $r\sim200$~pc (a typical scale over which
molecular gas and \ha\ emission are well correlated in galaxies) and
$r\sim1$~kpc (a typical scale for extragalactic studies) we measure
$\tau_{\rm dep}^{\rm mol}\approx1.6$~Gyr. The molecular depletion time
for the SMC as a whole is shorter, $\tau_{\rm dep}^{\rm
mol}\sim0.6$~Gyr, due to a large component of extended low level \ha\
emission.  These results are consistent with the typical depletion
time of $\tau_{\rm dep}^{\rm mol}\sim2$ Gyr observed in normal disks
on kpc scales
\citep{BIGIEL08,BIGIEL11}. Consequently the relation between
molecular gas and star formation activity appears to be at most only
weakly dependent on metallicity. This finding suggests that the
molecular content can be used to infer the star formation activity
(and viceversa) even in galaxies that are chemically primitive and
deficient in heavy elements.

We also measure the relation between star formation and total gas
surface density, which is dominated by \hi\ over most of the SMC
(Figure \ref{fig:sflgas}). We find that the relation is similarly
steep in the SMC ($1+p=2.2\pm0.1$) and in the outer disks of normal
galaxies \citep[$1+p\approx2$; ][]{BIGIEL10b}, but it is displaced toward
much larger surface densities. In the SMC the \hi\ surface density
reaches values as high as $\Shi\sim100$ \msunperpcsq, while in most
galaxies $\Shi\lesssim10$ \msunperpcsq. As a consequence the use of
the standard total gas to star formation rate relation dramatically
overpredicts the star formation activity over much of this source.
This finding supports the explanation that DLAs are deficient in star
formation because of very low molecular fractions \citep{WOLFE06}.
The gas (or \hi) depletion time for the SMC is approximately a Hubble
time, $\tau_{\rm dep}^{\rm gas}\approx11.8$~Gyr.

We compare our results with the recent analytical models by
\citet[][KMT09]{KRUMHOLZ09} and \citet[][OML10]{OSTRIKER10}.
For high-resolution observations, we find that KMT09 is very
successful at reproducing the displacement in the relation between
\Sgas\ and \Ssfr\ toward higher surface densities (Figure
\ref{fig:fig3}). We also find that with a clumping factor $c$ of unity
it approximately predicts the observed relation between the molecular
to atomic ratio $\Shtwo/\Shi$ and \Sgas\ (Figure \ref{fig:fig5}),
although on a linear scale it becomes apparent that there is a
systematic overprediction at high \Sgas\ by $0.3-0.5$ dex (Figure
\ref{fig:kmtcomp}).  The slope of \Ssfr\ vs. \Sgas\ predicted by KMT09
for fixed $cZ$ is steeper than the relation observed in the SMC, both
at the observed 12 pc resolution and when the data is averaged over
200 pc and 1 kpc scales (Figures \ref{fig:fig3} and Figure
\ref{fig:fig6}).  Given that KMT09 co-locates \hi\ and \htwo\ within
the same dense cloud complexes, a strong relation between \Shi\ and
\Ssfr\ would be expected at 200 pc scales, but this is not seen in the
observations (Figure \ref{fig:fig4}).  We test the effect of smoothing
the observations to larger spatial scales, and find that changing the
clumping parameter in the KMT09 model does not match the effect of
smoothing the observations (Figure \ref{fig:fig6}). The bulk of the
total gas surface densities in the SMC stay approximately constant for
spatial scales in the $r\sim12$~pc---1~kpc range.  This suggests that
much of the \hi\ is in a warm, diffuse component, which stands in
contrast to the starting assumption of KMT09 that the \hi\ consists
primarily of cold gas concentrated in the envelopes of dense
atomic-molecular complexes, with the warm \hi\ a small fraction of the
total ISM mass.

In comparing the observations to the predictions of OML10, we find
that the expected \Sgbc/\Sdiff\ is very large in comparison to the
observed \Shtwo/\Shi\ in the SMC (Figure \ref{fig:fig5}), if we adopt
the same heating efficiency and turbulence parameters for
thermal/dynamical equilibrium as in typical large spirals with $Z\sim
1$. In other words, the model conspicuosly overpredicts the molecular
fraction in the SMC for the default parameters. This motivates us to
investigate variations and extensions of the OML10 model.  In the
first variation, which we call OML10h, we introduce a metallicity
dependence in the ratio of gas heating to star formation that produces
enhanced heating at low metallicities.  This modification is based on
the idea that a lower dust-to-gas ratio allows UV to escape and travel
farther from star-forming regions, potentially raising $G_0'$
significantly in the diffuse phase.  An enhanced $G_0'$ leads to
increased heating in the diffuse medium, permitting thermal
equilibrium of warm \hi\ at high pressure.  Without enhanced heating,
warm gas at the typical thermal pressure in the SMC ($P_{\rm th}/k\sim
3\times 10^4$~K~cm$^{-3}$ for \Shi$\sim 65$ \msunperpcsq) would cool
very quickly.  Enhanced heating changes the balance between the
diffuse and self-gravitating phases of the gas.  As a consequence the
self-gravitating phase is less abundant than at normal metallicities
under similar conditions for \Sgas. This model does a very reasonable
job at matching the $\Shtwo/\Shi$ derived for the SMC (Figures
\ref{fig:fig5}), although there are still local discrepancies at the
level of $\sim\pm0.3$ dex.

As an extension of OML10, which we term OML10p, we do not change the
balance between the diffuse and self-gravitating components, but add
an accounting for the abundance of \hi\ within gravitationally bound
clouds, based on the photodissociation formalism of KMT09.  The \hi\
is split between the diffuse and the self-gravitating phases, and
under the typical SMC conditions, half of the \hi\ would be
self-gravitating.  The photodissociation calculation for the GBC
component can also be applied to the enhanced-heating model; we denote
this as model OML10ph.  Model OML10ph shares the reduced abundance of
the self-gravitating phase with model OML10h, and includes a
contribution to \hi\ that originates in cold self-gravitating clouds.
Note that in adding the photodissociation estimate for GBCs, we do not
change the predicted SFR.  Thus, model OML10 and OML10p have the same
\Ssfr\ vs. \Sgas\ as each other, as do model OML10h and OML10ph
(Figure \ref{fig:fig11}).

When compared to the observed \Shtwo/\Shi\ in the SMC, model OML10p
slightly overpredicts the molecular ratio ($\sim 10-30\%$
for the model vs. $\sim 5-15\%$ for the data; see Figure
\ref{fig:fig10}).  For the typical parameters of the SMC, $\sim
30-60\%$ of the total \hi\ would be in GBCs for model OML10p (Figure
\ref{fig:fig9}).  Although this is less than in KMT09 (which puts all
of the \hi\ in the envelopes of cold, dense cloud complexes), to some
extent it shares the difficulty that in observations \hi\ is not
correlated with star formation activity (Figure \ref{fig:fig4}).  Model
OML10ph, with a lower abundance of self-gravitating gas and hence
\Shtwo\ due to enhanced heating, follows the observed
\Shtwo/\Shi\ magnitude and slope better than model OML10p (Figure
\ref{fig:fig10}).  OML10ph also has a lower ratio of cold to warm
\hi\ than OML10p or KMT09, consistent with the observations of
\citet{DICKEY00}.  When compared to star formation in the SMC, the
enhanced-heating models (OML10h, OML10ph) fit the relation between
\Ssfr\ and \Sgas\ much better than the models that adopt the same
heating efficiency as higher-metallicity spirals (OML10, OML10p), as
shown in Figure \ref{fig:fig11}.

In essence, the observed star formation rate requires that the
eligible cold, self-gravitating gas (whether \hi\ or \htwo) is less
abundant at lower metallicities, otherwise we would expect larger star
formation rates for the observed \Sgas\ surface densities.  The
diminished abundance of the cold self-gravitating phase is supported
by observations of the distribution of \hi\ temperatures
\citep{DICKEY00}, and by the relative uniformity of \Shi\ when
averaged at different scales.  The possibility of an enhanced
radiation field that could maintain warm, diffuse \hi\ at high surface
density and pressure is suggested by recent dust emission modeling
\citep{SANDSTROM10}. Further observational confirmation of enhanced
heating could come from observations of the cooling in the diffuse
atomic gas. Further characterization of the fraction of cold atomic
gas and its spatial distribution in both Magellanic Clouds using \hi\
absorption/emission techniques would also be extremely valuable in
constraining the models.

\acknowledgments We thank Chris McKee, Mark Krumholz, Andrey Kravtsov, 
and Nick Gnedin for enlightening and stimulating discussions at
several stages of this research, and Jean-Philippe Bernard as well as
the anonymous referee for their comments on the manuscript.  A. B. and
K. J. wish to acknowledge partial support from grants NASA
JPL-1314022, NSF AST-0838178, NSF AST-0955836, as well as a Cottrell
Scholar award from the Research Corporation for Science
Advancement. E.C.O. acknowledges support from grant NSF AST-0908185.
P.F.W. acknowledges support from grant NSF AST-098566.  M.R. wishes to
acknowledge support from FONDECYT(CHILE) grant 1080335 and is
supported by the Chilean {\em Center for Astrophysics} FONDAP
15010003.  A.B. also acknowledges travel support from FONDECYT(CHILE)
grant 1080335.  This research has made use of NASA's Astrophysics Data
System.


\bibliographystyle{apj-hacked}

\bibliography{nsflib-proc} 

\begin{thebibliography}{115}
\expandafter\ifx\csname natexlab\endcsname\relax\def\natexlab#1{#1}\fi

\bibitem[{{Bekki} \& {Stanimirovi{\'c}}(2009)}]{BEKKI09}
{Bekki}, K., \& {Stanimirovi{\'c}}, S. 2009, \mnras, 395, 342

\bibitem[{{Bernard} {et~al.}(1999){Bernard}, {Abergel}, {Ristorcelli}, {Pajot},
  {Torre}, {Boulanger}, {Giard}, {Lagache}, {Serra}, {Lamarre}, {Puget},
  {Lepeintre}, \& {Cambr{\'e}sy}}]{BERNARD99}
{Bernard}, J.~P., {Abergel}, A., {Ristorcelli}, I., {et~al.} 1999, \aap, 347,
  640

\bibitem[{{Bigiel} {et~al.}(2010{\natexlab{a}}){Bigiel}, {Bolatto}, {Leroy},
  {Blitz}, {Walter}, {Rosolowsky}, {Lopez}, \& {Plambeck}}]{BIGIEL10a}
{Bigiel}, F., {Bolatto}, A.~D., {Leroy}, A.~K., {et~al.} 2010{\natexlab{a}},
  \apj, 725, 1159

\bibitem[{{Bigiel} {et~al.}(2010{\natexlab{b}}){Bigiel}, {Leroy}, {Walter},
  {Blitz}, {Brinks}, {de Blok}, \& {Madore}}]{BIGIEL10b}
{Bigiel}, F., {Leroy}, A., {Walter}, F., {et~al.} 2010{\natexlab{b}}, \aj, 140,
  1194

\bibitem[{{Bigiel} {et~al.}(2008){Bigiel}, {Leroy}, {Walter}, {Brinks}, {de
  Blok}, {Madore}, \& {Thornley}}]{BIGIEL08}
{Bigiel}, F., {Leroy}, A., ---. 2008, \aj, 136, 2846

\bibitem[{{Bigiel} {et~al.}(2011){Bigiel}, {Leroy}, {Walter}, {Brinks}, {de
  Blok}, {Kramer}, {Rix}, {Schruba}, {Schuster}, {Usero}, \&
  {Wiesemeyer}}]{BIGIEL11}
{Bigiel}, F., {Leroy}, A.~K., ---. 2011, ArXiv e-prints

\bibitem[{{Blanc} {et~al.}(2009){Blanc}, {Heiderman}, {Gebhardt}, {Evans}, \&
  {Adams}}]{BLANC09}
{Blanc}, G.~A., {Heiderman}, A., {Gebhardt}, K., {Evans}, N.~J., \& {Adams}, J.
  2009, \apj, 704, 842

\bibitem[{{Blitz} {et~al.}(2007){Blitz}, {Fukui}, {Kawamura}, {Leroy},
  {Mizuno}, \& {Rosolowsky}}]{BLITZ07}
{Blitz}, L., {Fukui}, Y., {Kawamura}, A., {et~al.} 2007, Protostars and Planets
  V, 81

\bibitem[{{Blitz} \& {Rosolowsky}(2006)}]{BLITZ06}
{Blitz}, L., \& {Rosolowsky}, E. 2006, \apj, 650, 933

\bibitem[{{Bloemen} {et~al.}(1990){Bloemen}, {Deul}, \& {Thaddeus}}]{BLOEMEN90}
{Bloemen}, J.~B.~G.~M., {Deul}, E.~R., \& {Thaddeus}, P. 1990, \aap, 233, 437

\bibitem[{{Boissier} {et~al.}(2003){Boissier}, {Prantzos}, {Boselli}, \&
  {Gavazzi}}]{BOISSIER03}
{Boissier}, S., {Prantzos}, N., {Boselli}, A., \& {Gavazzi}, G. 2003, \mnras,
  346, 1215

\bibitem[{{Bolatto} {et~al.}(1999){Bolatto}, {Jackson}, \&
  {Ingalls}}]{BOLATTO99}
{Bolatto}, A.~D., {Jackson}, J.~M., \& {Ingalls}, J.~G. 1999, \apj, 513, 275

\bibitem[{{Bolatto} {et~al.}(2008){Bolatto}, {Leroy}, {Rosolowsky}, {Walter},
  \& {Blitz}}]{BOLATTO08}
{Bolatto}, A.~D., {Leroy}, A.~K., {Rosolowsky}, E., {Walter}, F., \& {Blitz},
  L. 2008, \apj, 686, 948

\bibitem[{{Bolatto} {et~al.}(2007){Bolatto}, {Simon}, {Stanimirovi{\'c}}, {van
  Loon}, {Shah}, {Venn}, {Leroy}, {Sandstrom}, {Jackson}, {Israel}, {Li},
  {Staveley-Smith}, {Bot}, {Boulanger}, \& {Rubio}}]{BOLATTO07}
{Bolatto}, A.~D., {Simon}, J.~D., {Stanimirovi{\'c}}, S., {et~al.} 2007, \apj,
  655, 212

\bibitem[{{Bot} {et~al.}(2004){Bot}, {Boulanger}, {Lagache}, {Cambr{\'e}sy}, \&
  {Egret}}]{BOT04}
{Bot}, C., {Boulanger}, F., {Lagache}, G., {Cambr{\'e}sy}, L., \& {Egret}, D.
  2004, \aap, 423, 567

\bibitem[{{Boulanger} {et~al.}(1996){Boulanger}, {Abergel}, {Bernard},
  {Burton}, {Desert}, {Hartmann}, {Lagache}, \& {Puget}}]{BOULANGER96}
{Boulanger}, F., {Abergel}, A., {Bernard}, J., {et~al.} 1996, \aap, 312, 256

\bibitem[{{Burkhart} {et~al.}(2010){Burkhart}, {Stanimirovi{\'c}}, {Lazarian},
  \& {Kowal}}]{BURKHART10}
{Burkhart}, B., {Stanimirovi{\'c}}, S., {Lazarian}, A., \& {Kowal}, G. 2010,
  \apj, 708, 1204

\bibitem[{{Calzetti} {et~al.}(2007){Calzetti}, {Kennicutt}, {Engelbracht},
  {Leitherer}, {Draine}, {Kewley}, {Moustakas}, {Sosey}, {Dale}, {Gordon},
  {Helou}, {Hollenbach}, {Armus}, {Bendo}, {Bot}, {Buckalew}, {Jarrett}, {Li},
  {Meyer}, {Murphy}, {Prescott}, {Regan}, {Rieke}, {Roussel}, {Sheth}, {Smith},
  {Thornley}, \& {Walter}}]{CALZETTI07}
{Calzetti}, D., {Kennicutt}, R.~C., {Engelbracht}, C.~W., {et~al.} 2007, \apj,
  666, 870

\bibitem[{{Caplan} {et~al.}(1996){Caplan}, {Ye}, {Deharveng}, {Turtle}, \&
  {Kennicutt}}]{CAPLAN96}
{Caplan}, J., {Ye}, T., {Deharveng}, L., {Turtle}, A.~J., \& {Kennicutt}, R.~C.
  1996, \aap, 307, 403

\bibitem[{{Dame} {et~al.}(2001){Dame}, {Hartmann}, \& {Thaddeus}}]{DAME01}
{Dame}, T.~M., {Hartmann}, D., \& {Thaddeus}, P. 2001, \apj, 547, 792

\bibitem[{{Dickey} {et~al.}(2000){Dickey}, {Mebold}, {Stanimirovi{\'c}}, \&
  {Staveley-Smith}}]{DICKEY00}
{Dickey}, J.~M., {Mebold}, U., {Stanimirovi{\'c}}, S., \& {Staveley-Smith}, L.
  2000, \apj, 536, 756

\bibitem[{{Dobashi} {et~al.}(2009){Dobashi}, {Bernard}, {Kawamura}, {Egusa},
  {Hughes}, {Paradis}, {Bot}, \& {Reach}}]{DOBASHI09}
{Dobashi}, K., {Bernard}, J., {Kawamura}, A., {et~al.} 2009, \aj, 137, 5099

\bibitem[{{Draine} \& {Li}(2001)}]{DRAINE01}
{Draine}, B.~T., \& {Li}, A. 2001, \apj, 551, 807

\bibitem[{{Draine} {et~al.}(2007){Draine}, {Dale}, {Bendo}, {Gordon}, {Smith},
  {Armus}, {Engelbracht}, {Helou}, {Kennicutt}, {Li}, {Roussel}, {Walter},
  {Calzetti}, {Moustakas}, {Murphy}, {Rieke}, {Bot}, {Hollenbach}, {Sheth}, \&
  {Teplitz}}]{DRAINE07}
{Draine}, B.~T., {Dale}, D.~A., {Bendo}, G., {et~al.} 2007, \apj, 663, 866

\bibitem[{{Dufour}(1984)}]{DUFOUR84}
{Dufour}, R.~J. 1984, in IAU Symposium, Vol. 108, Structure and Evolution of
  the Magellanic Clouds, ed. {S.~van den Bergh \& K.~S.~D.~de Boer}, 353--360

\bibitem[{{Engelbracht} {et~al.}(2008){Engelbracht}, {Rieke}, {Gordon},
  {Smith}, {Werner}, {Moustakas}, {Willmer}, \& {Vanzi}}]{ENGELBRACHT08}
{Engelbracht}, C.~W., {Rieke}, G.~H., {Gordon}, K.~D., {et~al.} 2008, \apj,
  678, 804

\bibitem[{{Flagey} {et~al.}(2009){Flagey}, {Noriega-Crespo}, {Boulanger},
  {Carey}, {Brooke}, {Falgarone}, {Huard}, {McCabe}, {Miville-Desch{\^e}nes},
  {Padgett}, {Paladini}, \& {Rebull}}]{FLAGEY09}
{Flagey}, N., {Noriega-Crespo}, A., {Boulanger}, F., {et~al.} 2009, \apj, 701,
  1450

\bibitem[{{Fukui} {et~al.}(1999){Fukui}, {Mizuno}, {Yamaguchi}, {Mizuno},
  {Onishi}, {Ogawa}, {Yonekura}, {Kawamura}, {Tachihara}, {Xiao}, {Yamaguchi},
  {Hara}, {Hayakawa}, {Kato}, {Abe}, {Saito}, {Mano}, {Matsunaga}, {Mine},
  {Moriguchi}, {Aoyama}, {Asayama}, {Yoshikawa}, \& {Rubio}}]{FUKUI99}
{Fukui}, Y., {Mizuno}, N., {Yamaguchi}, R., {et~al.} 1999, \pasj, 51, 745

\bibitem[{{Gardan} {et~al.}(2007){Gardan}, {Braine}, {Schuster}, {Brouillet},
  \& {Sievers}}]{GARDAN07}
{Gardan}, E., {Braine}, J., {Schuster}, K.~F., {Brouillet}, N., \& {Sievers},
  A. 2007, \aap, 473, 91

\bibitem[{{Glover} \& {Mac Low}(2010)}]{GLOVER10}
{Glover}, S.~C.~O., \& {Mac Low}, M. 2010, ArXiv e-prints

\bibitem[{{Gnedin} \& {Kravtsov}(2010{\natexlab{a}})}]{GNEDIN10b}
{Gnedin}, N.~Y., \& {Kravtsov}, A.~V. 2010{\natexlab{a}}, ArXiv e-prints

\bibitem[{{Gnedin} \& {Kravtsov}(2010{\natexlab{b}})}]{GNEDIN10a}
---. 2010{\natexlab{b}}, \apj, 714, 287

\bibitem[{{Gordon} {et~al.}(2011){Gordon}, {Meixner}, \& {SAGE-SMC
  team}}]{GORDON11}
{Gordon}, K., {Meixner}, M., \& {SAGE-SMC team}. 2011, \apj, in prep.

\bibitem[{{Harris} \& {Zaritsky}(2004)}]{HARRIS04}
{Harris}, J., \& {Zaritsky}, D. 2004, \aj, 127, 1531

\bibitem[{{Haschke} {et~al.}(2011){Haschke}, {Grebel}, \& {Duffau}}]{HASCHKE11}
{Haschke}, R., {Grebel}, E.~K., \& {Duffau}, S. 2011, \aj, 141, 158

\bibitem[{{Heiles} \& {Troland}(2003)}]{HEILES03}
{Heiles}, C., \& {Troland}, T.~H. 2003, \apj, 586, 1067

\bibitem[{{Heyer} {et~al.}(2009){Heyer}, {Krawczyk}, {Duval}, \&
  {Jackson}}]{HEYER09}
{Heyer}, M., {Krawczyk}, C., {Duval}, J., \& {Jackson}, J.~M. 2009, \apj, 699,
  1092

\bibitem[{{Hilditch} {et~al.}(2005){Hilditch}, {Howarth}, \&
  {Harries}}]{HILDITCH05}
{Hilditch}, R.~W., {Howarth}, I.~D., \& {Harries}, T.~J. 2005, \mnras, 357, 304

\bibitem[{{Hoopes} {et~al.}(1999){Hoopes}, {Walterbos}, \& {Rand}}]{HOOPES99}
{Hoopes}, C.~G., {Walterbos}, R.~A.~M., \& {Rand}, R.~J. 1999, \apj, 522, 669

\bibitem[{{Hughes} {et~al.}(2010){Hughes}, {Wong}, {Ott}, {M\"uller}, {Pineda},
  {Mizuno}, {Bernard}, {Paradis}, {Maddison}, {Reach}, {Staveley-Smith},
  {Kawamura}, {Meixner}, {Kim}, {Onishi}, {Mizuno}, \& {Fukui}}]{HUGHES10}
{Hughes}, A., {Wong}, T., {Ott}, J., {et~al.} 2010, \mnras, 406, 2065

\bibitem[{{Israel}(1997)}]{ISRAEL97}
{Israel}, F.~P. 1997, \aap, 328, 471

\bibitem[{{Israel} {et~al.}(1986){Israel}, {de Graauw}, {van de Stadt}, \& {de
  Vries}}]{ISRAEL86}
{Israel}, F.~P., {de Graauw}, T., {van de Stadt}, H., \& {de Vries}, C.~P.
  1986, \apj, 303, 186

\bibitem[{{Israel} {et~al.}(1996){Israel}, {Maloney}, {Geis}, {Herrmann},
  {Madden}, {Poglitsch}, \& {Stacey}}]{ISRAEL96}
{Israel}, F.~P., {Maloney}, P.~R., {Geis}, N., {et~al.} 1996, \apj, 465, 738

\bibitem[{{Israel} {et~al.}(1993){Israel}, {Johansson}, {Lequeux}, {Booth},
  {Nyman}, {Crane}, {Rubio}, {de Graauw}, {Kutner}, {Gredel}, {Boulanger},
  {Garay}, \& {Westerlund}}]{ISRAEL93}
{Israel}, F.~P., {Johansson}, L.~E.~B., {Lequeux}, J., {et~al.} 1993, \aap,
  276, 25

\bibitem[{{Israel} {et~al.}(2003){Israel}, {Johansson}, {Rubio}, {Garay}, {de
  Graauw}, {Booth}, {Boulanger}, {Kutner}, {Lequeux}, \& {Nyman}}]{ISRAEL03}
{Israel}, F.~P., {Johansson}, L.~E.~B., {Rubio}, M., {et~al.} 2003, \aap, 406,
  817

\bibitem[{{Keller} \& {Wood}(2006)}]{KELLER06}
{Keller}, S.~C., \& {Wood}, P.~R. 2006, \apj, 642, 834

\bibitem[{{Kennicutt}(1989)}]{KENNICUTT89}
{Kennicutt}, Jr., R.~C. 1989, \apj, 344, 685

\bibitem[{{Kennicutt}(1998)}]{KENNICUTT98}
---. 1998, \apj, 498, 541

\bibitem[{{Kennicutt} {et~al.}(1995){Kennicutt}, {Bresolin}, {Bomans},
  {Bothun}, \& {Thompson}}]{KENNICUTT95}
{Kennicutt}, Jr., R.~C., {Bresolin}, F., {Bomans}, D.~J., {Bothun}, G.~D., \&
  {Thompson}, I.~B. 1995, \aj, 109, 594

\bibitem[{{Krumholz} {et~al.}(2009{\natexlab{a}}){Krumholz}, {Ellison},
  {Prochaska}, \& {Tumlinson}}]{KRUMHOLZ09b}
{Krumholz}, M.~R., {Ellison}, S.~L., {Prochaska}, J.~X., \& {Tumlinson}, J.
  2009{\natexlab{a}}, \apjl, 701, L12

\bibitem[{{Krumholz} {et~al.}(2011){Krumholz}, {Leroy}, \&
  {McKee}}]{KRUMHOLZ11}
{Krumholz}, M.~R., {Leroy}, A.~K., \& {McKee}, C.~F. 2011, ArXiv e-prints

\bibitem[{{Krumholz} \& {McKee}(2005)}]{KRUMHOLZ05}
{Krumholz}, M.~R., \& {McKee}, C.~F. 2005, \apj, 630, 250

\bibitem[{{Krumholz} {et~al.}(2009{\natexlab{b}}){Krumholz}, {McKee}, \&
  {Tumlinson}}]{KRUMHOLZ09c}
{Krumholz}, M.~R., {McKee}, C.~F., \& {Tumlinson}, J. 2009{\natexlab{b}}, \apj,
  693, 216

\bibitem[{{Krumholz} {et~al.}(2009{\natexlab{c}}){Krumholz}, {McKee}, \&
  {Tumlinson}}]{KRUMHOLZ09}
---. 2009{\natexlab{c}}, \apj, 699, 850

\bibitem[{{Kurt} {et~al.}(1999){Kurt}, {Dufour}, {Garnett}, {Skillman},
  {Mathis}, {Peimbert}, {Torres-Peimbert}, \& {Ruiz}}]{KURT99}
{Kurt}, C.~M., {Dufour}, R.~J., {Garnett}, D.~R., {et~al.} 1999, \apj, 518, 246

\bibitem[{{Lada} {et~al.}(2009){Lada}, {Lombardi}, \& {Alves}}]{LADA09}
{Lada}, C.~J., {Lombardi}, M., \& {Alves}, J.~F. 2009, \apj, 703, 52

\bibitem[{{Leitherer} \& {Heckman}(1995)}]{LEITHERER95}
{Leitherer}, C., \& {Heckman}, T.~M. 1995, \apjs, 96, 9

\bibitem[{{Lequeux} {et~al.}(1994){Lequeux}, {Le Bourlot}, {Pineau des Forets},
  {Roueff}, {Boulanger}, \& {Rubio}}]{LEQUEUX94}
{Lequeux}, J., {Le Bourlot}, J., {Pineau des Forets}, G., {et~al.} 1994, \aap,
  292, 371

\bibitem[{{Leroy} {et~al.}(2007{\natexlab{a}}){Leroy}, {Bolatto},
  {Stanimirovi{\'c}}, {Mizuno}, {Israel}, \& {Bot}}]{LEROY07}
{Leroy}, A., {Bolatto}, A., {Stanimirovi{\'c}}, S., {et~al.}
  2007{\natexlab{a}}, \apj, 658, 1027

\bibitem[{{Leroy} {et~al.}(2007{\natexlab{b}}){Leroy}, {Cannon}, {Walter},
  {Bolatto}, \& {Weiss}}]{LEROY07B}
{Leroy}, A., {Cannon}, J., {Walter}, F., {Bolatto}, A., \& {Weiss}, A.
  2007{\natexlab{b}}, \apj, 663, 990

\bibitem[{{Leroy} {et~al.}(2008){Leroy}, {Walter}, {Brinks}, {Bigiel}, {de
  Blok}, {Madore}, \& {Thornley}}]{LEROY08}
{Leroy}, A.~K., {Walter}, F., {Brinks}, E., {et~al.} 2008, \aj, 136, 2782

\bibitem[{{Leroy} {et~al.}(2009){Leroy}, {Bolatto}, {Bot}, {Engelbracht},
  {Gordon}, {Israel}, {Rubio}, {Sandstrom}, \& {Stanimirovi{\'c}}}]{LEROY09}
{Leroy}, A.~K., {Bolatto}, A., {Bot}, C., {et~al.} 2009, \apj, 702, 352

\bibitem[{{Leroy} {et~al.}(2011){Leroy}, {Bolatto}, {Gordon}, {Sandstrom},
  {Gratier}, {Rosolowsky}, {Engelbracht}, {Mizuno}, {Corbelli}, {Fukui}, \&
  {Kawamura}}]{LEROY11}
{Leroy}, A.~K., {Bolatto}, A.~D., {Gordon}, K., {et~al.} 2011, \apj, in press

\bibitem[{{Li} \& {Draine}(2002)}]{LI02}
{Li}, A., \& {Draine}, B.~T. 2002, \apj, 576, 762

\bibitem[{{Madden} {et~al.}(2006){Madden}, {Galliano}, {Jones}, \&
  {Sauvage}}]{MADDEN06}
{Madden}, S.~C., {Galliano}, F., {Jones}, A.~P., \& {Sauvage}, M. 2006, \aap,
  446, 877

\bibitem[{{Madden} {et~al.}(1997){Madden}, {Poglitsch}, {Geis}, {Stacey}, \&
  {Townes}}]{MADDEN97}
{Madden}, S.~C., {Poglitsch}, A., {Geis}, N., {Stacey}, G.~J., \& {Townes},
  C.~H. 1997, \apj, 483, 200

\bibitem[{{Maloney} \& {Black}(1988)}]{MALONEY88}
{Maloney}, P., \& {Black}, J.~H. 1988, \apj, 325, 389

\bibitem[{{Martin} \& {Kennicutt}(2001)}]{MARTIN01}
{Martin}, C.~L., \& {Kennicutt}, Jr., R.~C. 2001, \apj, 555, 301

\bibitem[{{Mathis} {et~al.}(1983){Mathis}, {Mezger}, \& {Panagia}}]{MATHIS83}
{Mathis}, J.~S., {Mezger}, P.~G., \& {Panagia}, N. 1983, \aap, 128, 212

\bibitem[{{McGee} \& {Newton}(1981)}]{MCGEE81}
{McGee}, R.~X., \& {Newton}, L.~M. 1981, Proceedings of the Astronomical
  Society of Australia, 4, 189

\bibitem[{{McKee}(1989)}]{MCKEE89}
{McKee}, C.~F. 1989, \apj, 345, 782

\bibitem[{{McKee} \& {Ostriker}(2007)}]{MO07}
{McKee}, C.~F., \& {Ostriker}, E.~C. 2007, \araa, 45, 565

\bibitem[{{Mizuno} {et~al.}(2001){Mizuno}, {Rubio}, {Mizuno}, {Yamaguchi},
  {Onishi}, \& {Fukui}}]{MIZUNO01}
{Mizuno}, N., {Rubio}, M., {Mizuno}, A., {et~al.} 2001, \pasj, 53, L45

\bibitem[{{M\"uller} {et~al.}(2010){M\"uller}, {Ott}, {Hughes}, {Pineda},
  {Wong}, {Mizuno}, {Kawamura}, {Mizuno}, {Fukui}, {Onishi}, \&
  {Rubio}}]{MULLER10}
{M\"uller}, E., {Ott}, J., {Hughes}, A., {et~al.} 2010, \apj, 712, 1248

\bibitem[{{Nieten} {et~al.}(2006){Nieten}, {Neininger}, {Gu{\'e}lin},
  {Ungerechts}, {Lucas}, {Berkhuijsen}, {Beck}, \& {Wielebinski}}]{NIETEN06}
{Nieten}, C., {Neininger}, N., {Gu{\'e}lin}, M., {et~al.} 2006, \aap, 453, 459

\bibitem[{{Oliveira} {et~al.}(2011){Oliveira}, {van Loon}, {Sloan},
  {Indebetouw}, {Kemper}, {Tielens}, {Simon}, {Woods}, \&
  {Meixner}}]{OLIVEIRA11}
{Oliveira}, J.~M., {van Loon}, J.~T., {Sloan}, G.~C., {et~al.} 2011, \mnras,
  411, L36

\bibitem[{{Ostriker} {et~al.}(2010){Ostriker}, {McKee}, \&
  {Leroy}}]{OSTRIKER10}
{Ostriker}, E.~C., {McKee}, C.~F., \& {Leroy}, A.~K. 2010, \apj, 721, 975

\bibitem[{{Pagel}(2003)}]{PAGEL03}
{Pagel}, B.~E.~J. 2003, in Astronomical Society of the Pacific Conference
  Series, Vol. 304, Astronomical Society of the Pacific Conference Series, ed.
  {C.~Charbonnel, D.~Schaerer, \& G.~Meynet}, 187--+

\bibitem[{{Planck Collaboration} {et~al.}(2011{\natexlab{a}}){Planck
  Collaboration}, {Ade}, {Aghanim}, {Arnaud}, {Ashdown}, {Aumont},
  {Baccigalupi}, {Balbi}, {Banday}, {Barreiro}, {Bartlett}, {Battaner},
  {Benabed}, {Beno{\^i}t}, {Bernard}, {Bersanelli}, {Bhatia}, {Bock},
  {Bonaldi}, {Bond}, {Borrill}, {Bot}, {Bouchet}, {Boulanger}, {Bucher},
  {Burigana}, {Cabella}, {Cardoso}, {Catalano}, {Cay{\'o}n}, {Challinor},
  {Chamballu}, {Chiang}, {Chiang}, {Christensen}, {Clements}, {Colombi},
  {Couchot}, {Coulais}, {Crill}, {Cuttaia}, {Danese}, {Davies}, {Davis}, {de
  Bernardis}, {de Gasperis}, {de Rosa}, {de Zotti}, {Delabrouille}, {Delouis},
  {D{\'e}sert}, {Dickinson}, {Dobashi}, {Doerl}, {Donzelli}, {Dor{\'e}},
  {Douspis}, {Dupac}, {Efstathiou}, {En$\backslash$sslin}, {Finelli}, {Forni},
  {Frailis}, {Franceschi}, {Fukui}, {Galeotta}, {Ganga}, {Giard}, {Giardino},
  {Giraud-H{\'e}raud}, {Gonz{\'a}lez-Nuevo}, {G{\'o}rski}, {Gratton},
  {Gregorio}, {Gruppuso}, {Harrison}, {Helou}, {Henrot-Versill{\'e}},
  {Herranz}, {Hildebrandt}, {Hivon}, {Hobson}, {Holmes}, {Hovest}, {Hoyland},
  {Huffenberger}, {Jaffe}, {Jones}, {Juvela}, {Kawamura}, {Keih{\"a}nen},
  {Keskitalo}, {Kisner}, {Kneissl}, {Knox}, {Kurki-Suonio}, {Lagache},
  {L{\"a}hteenm{\"a}ki}, {Lamarre}, {Lasenby}, {Laureijs}, {Lawrence}, {Leach},
  {Leonardi}, {Leroy}, {Linden-V$\backslash$ornle}, {L{\'o}pez-Caniego},
  {Lubin}, {Mac{\'{\i}}as-P{\'e}rez}, {MacTavish}, {Madden}, {Maffei},
  {Mandolesi}, {Mann}, {Maris}, {Mart{\'{\i}}nez-Gonz{\'a}lez}, {Masi},
  {Matarrese}, {Matthai}, {Mazzotta}, {Meinhold}, {Melchiorri}, {Mendes},
  {Mennella}, {Miville-Desch{\^e}nes}, {Moneti}, {Montier}, {Morgante},
  {Mortlock}, {Munshi}, {Murphy}, {Naselsky}, {Nati}, {Natoli}, {Netterfield},
  {N$\backslash$orgaard-Nielsen}, {Noviello}, {Novikov}, {Novikov}, {Onishi},
  {Osborne}, {Pajot}, {Paladini}, {Paradis}, {Pasian}, {Patanchon},
  {Perdereau}, {Perotto}, {Perrotta}, {Piacentini}, {Piat}, {Plaszczynski},
  {Pointecouteau}, {Polenta}, {Ponthieu}, {Poutanen}, {Pr{\'e}zeau}, {Prunet},
  {Puget}, {Reach}, {Rebolo}, {Reinecke}, {Renault}, {Ricciardi}, {Riller},
  {Ristorcelli}, {Rocha}, {Rosset}, {Rowan-Robinson},
  {Rubi{\~n}o-Mart{\'{\i}}n}, {Rusholme}, {Sandri}, {Savini}, {Scott},
  {Seiffert}, {Smoot}, {Starck}, {Stivoli}, {Stolyarov}, {Sudiwala}, {Sygnet},
  {Tauber}, {Terenzi}, {Toffolatti}, {Tomasi}, {Torre}, {Tristram}, {Tuovinen},
  {Umana}, {Valenziano}, {Varis}, {Vielva}, {Villa}, {Vittorio}, {Wade},
  {Wandelt}, {Ysard}, {Yvon}, {Zacchei}, \& {Zonca}}]{PLANCKCOLLABORATION11}
{Planck Collaboration}{Ade}, P.~A.~R., {Aghanim}, N., {et~al.}
  2011{\natexlab{a}}, ArXiv e-prints

\bibitem[{{Planck Collaboration} {et~al.}(2011{\natexlab{b}}){Planck
  Collaboration}, {Abergel}, {Ade}, {Aghanim}, {Arnaud}, {Ashdown}, {Aumont},
  {Baccigalupi}, {Balbi}, {\~{}J.~Banday}, \& et~al.}]{PLANCKTAURUS11}
{Planck Collaboration}{Abergel}, A., {Ade}, P., {et~al.} 2011{\natexlab{b}},
  ArXiv e-prints

\bibitem[{{Rahman} {et~al.}(2011){Rahman}, {Bolatto}, {Wong}, {Leroy},
  {Walter}, {Rosolowsky}, {West}, {Bigiel}, {Ott}, {Xue}, {Herrera-Camus},
  {Jameson}, {Blitz}, \& {Vogel}}]{RAHMAN11}
{Rahman}, N., {Bolatto}, A.~D., {Wong}, T., {et~al.} 2011, \apj, 730, 72

\bibitem[{{Rela{\~n}o} {et~al.}(2010){Rela{\~n}o}, {Monreal-Ibero},
  {V{\'{\i}}lchez}, \& {Kennicutt}}]{RELANO10}
{Rela{\~n}o}, M., {Monreal-Ibero}, A., {V{\'{\i}}lchez}, J.~M., \& {Kennicutt},
  R.~C. 2010, \mnras, 402, 1635

\bibitem[{{Robertson} \& {Kravtsov}(2008)}]{ROBERTSON08}
{Robertson}, B.~E., \& {Kravtsov}, A.~V. 2008, \apj, 680, 1083

\bibitem[{{Roychowdhury} {et~al.}(2009){Roychowdhury}, {Chengalur}, {Begum}, \&
  {Karachentsev}}]{ROYCHOWDHURY09}
{Roychowdhury}, S., {Chengalur}, J.~N., {Begum}, A., \& {Karachentsev}, I.~D.
  2009, \mnras, 397, 1435

\bibitem[{{Roychowdhury} {et~al.}(2011){Roychowdhury}, {Chengalur}, {Kaisin},
  {Begum}, \& {Karachentsev}}]{ROYCHOWDHURY11}
{Roychowdhury}, S., {Chengalur}, J.~N., {Kaisin}, S.~S., {Begum}, A., \&
  {Karachentsev}, I.~D. 2011, ArXiv e-prints

\bibitem[{{Rubin} {et~al.}(2009){Rubin}, {Hony}, {Madden}, {Tielens},
  {Meixner}, {Indebetouw}, {Reach}, {Ginsburg}, {Kim}, {Mochizuki}, {Babler},
  {Block}, {Bracker}, {Engelbracht}, {For}, {Gordon}, {Hora}, {Leitherer},
  {Meade}, {Misselt}, {Sewilo}, {Vijh}, \& {Whitney}}]{RUBIN09}
{Rubin}, D., {Hony}, S., {Madden}, S.~C., {et~al.} 2009, \aap, 494, 647

\bibitem[{{Rubio} {et~al.}(1993){Rubio}, {Lequeux}, \& {Boulanger}}]{RUBIO93b}
{Rubio}, M., {Lequeux}, J., \& {Boulanger}, F. 1993, \aap, 271, 9

\bibitem[{{Sandstrom} {et~al.}(2010){Sandstrom}, {Bolatto}, {Draine}, {Bot}, \&
  {Stanimirovi{\'c}}}]{SANDSTROM10}
{Sandstrom}, K.~M., {Bolatto}, A.~D., {Draine}, B.~T., {Bot}, C., \&
  {Stanimirovi{\'c}}, S. 2010, \apj, 715, 701

\bibitem[{{Sanduleak}(1969)}]{SANDULEAK69}
{Sanduleak}, N. 1969, \aj, 74, 47

\bibitem[{{Schaye}(2004)}]{SCHAYE04}
{Schaye}, J. 2004, \apj, 609, 667

\bibitem[{{Schaye} \& {Dalla Vecchia}(2008)}]{SCHAYE08}
{Schaye}, J., \& {Dalla Vecchia}, C. 2008, \mnras, 383, 1210

\bibitem[{{Schmidt}(1959)}]{SCHMIDT59}
{Schmidt}, M. 1959, \apj, 129, 243

\bibitem[{{Schnee} {et~al.}(2008){Schnee}, {Li}, {Goodman}, \&
  {Sargent}}]{SCHNEE08}
{Schnee}, S., {Li}, J., {Goodman}, A.~A., \& {Sargent}, A.~I. 2008, \apj, 684,
  1228

\bibitem[{{Schruba} {et~al.}(2010){Schruba}, {Leroy}, {Walter}, {Sandstrom}, \&
  {Rosolowsky}}]{SCHRUBA10}
{Schruba}, A., {Leroy}, A.~K., {Walter}, F., {Sandstrom}, K., \& {Rosolowsky},
  E. 2010, \apj, 722, 1699

\bibitem[{{Schruba} {et~al.}(2011){Schruba}, {Leroy}, {Walter}, {Bigiel},
  {Brinks}, {de Blok}, {Dumas}, {Kramer}, {Rosolowsky}, {Sandstrom},
  {Schuster}, {Usero}, {Weiss}, \& {Wiesemeyer}}]{SCHRUBA11}
{Schruba}, A., {Leroy}, A.~K., {Walter}, F., {et~al.} 2011, \aj, submitted

\bibitem[{{Smith} \& {The MCELS Team}(1999)}]{SMITH99}
{Smith}, R.~C., \& {The MCELS Team}. 1999, in IAU Symposium, Vol. 190, New
  Views of the Magellanic Clouds, ed. {Y.-H.~Chu, N.~Suntzeff, J.~Hesser, \&
  D.~Bohlender}, 28--+

\bibitem[{{Stacey} {et~al.}(1991){Stacey}, {Geis}, {Genzel}, {Lugten},
  {Poglitsch}, {Sternberg}, \& {Townes}}]{STACEY91}
{Stacey}, G.~J., {Geis}, N., {Genzel}, R., {et~al.} 1991, \apj, 373, 423

\bibitem[{{Stanimirovi{\'c}} {et~al.}(1999){Stanimirovi{\'c}},
  {Staveley-Smith}, {Dickey}, {Sault}, \& {Snowden}}]{STANIMIROVIC99}
{Stanimirovi{\'c}}, S., {Staveley-Smith}, L., {Dickey}, J.~M., {Sault}, R.~J.,
  \& {Snowden}, S.~L. 1999, \mnras, 302, 417

\bibitem[{{Stanimirovi{\'c}} {et~al.}(2004){Stanimirovi{\'c}},
  {Staveley-Smith}, \& {Jones}}]{STANIMIROVIC04}
{Stanimirovi{\'c}}, S., {Staveley-Smith}, L., \& {Jones}, P.~A. 2004, \apj,
  604, 176

\bibitem[{{Stanimirovi{\'c}} {et~al.}(2000){Stanimirovi{\'c}},
  {Staveley-Smith}, {van der Hulst}, {Bontekoe}, {Kester}, \&
  {Jones}}]{STANIMIROVIC00}
{Stanimirovi{\'c}}, S., {Staveley-Smith}, L., {van der Hulst}, J.~M., {et~al.}
  2000, \mnras, 315, 791

\bibitem[{{Szewczyk} {et~al.}(2009){Szewczyk}, {Pietrzy{\'n}ski}, {Gieren},
  {Ciechanowska}, {Bresolin}, \& {Kudritzki}}]{SZEWCZYK09}
{Szewczyk}, O., {Pietrzy{\'n}ski}, G., {Gieren}, W., {et~al.} 2009, \aj, 138,
  1661

\bibitem[{{Tan}(2000)}]{TAN00}
{Tan}, J.~C. 2000, \apj, 536, 173

\bibitem[{{Tumlinson} {et~al.}(2002){Tumlinson}, {Shull}, {Rachford},
  {Browning}, {Snow}, {Fullerton}, {Jenkins}, {Savage}, {Crowther}, {Moos},
  {Sembach}, {Sonneborn}, \& {York}}]{TUMLINSON02}
{Tumlinson}, J., {Shull}, J.~M., {Rachford}, B.~L., {et~al.} 2002, \apj, 566,
  857

\bibitem[{{van Loon} {et~al.}(2010){van Loon}, {Oliveira}, {Gordon}, {Sloan},
  \& {Engelbracht}}]{VANLOON10b}
{van Loon}, J.~T., {Oliveira}, J.~M., {Gordon}, K.~D., {Sloan}, G.~C., \&
  {Engelbracht}, C.~W. 2010, \aj, 139, 1553

\bibitem[{{Walter} {et~al.}(2007){Walter}, {Cannon}, {Roussel}, {Bendo},
  {Calzetti}, {Dale}, {Draine}, {Helou}, {Kennicutt}, {Moustakas}, {Rieke},
  {Armus}, {Engelbracht}, {Gordon}, {Hollenbach}, {Lee}, {Li}, {Meyer},
  {Murphy}, {Regan}, {Smith}, {Brinks}, {de Blok}, {Bigiel}, \&
  {Thornley}}]{WALTER07}
{Walter}, F., {Cannon}, J.~M., {Roussel}, H., {et~al.} 2007, \apj, 661, 102

\bibitem[{{Wilke} {et~al.}(2004){Wilke}, {Klaas}, {Lemke}, {Mattila},
  {Stickel}, \& {Haas}}]{WILKE04}
{Wilke}, K., {Klaas}, U., {Lemke}, D., {et~al.} 2004, \aap, 414, 69

\bibitem[{{Wolfe} \& {Chen}(2006)}]{WOLFE06}
{Wolfe}, A.~M., \& {Chen}, H. 2006, \apj, 652, 981

\bibitem[{{Wolfire} {et~al.}(2010){Wolfire}, {Hollenbach}, \&
  {McKee}}]{WOLFIRE10}
{Wolfire}, M.~G., {Hollenbach}, D., \& {McKee}, C.~F. 2010, \apj, 716, 1191

\bibitem[{{Wolfire} {et~al.}(1995){Wolfire}, {Hollenbach}, {McKee}, {Tielens},
  \& {Bakes}}]{WOLFIRE95}
{Wolfire}, M.~G., {Hollenbach}, D., {McKee}, C.~F., {Tielens}, A.~G.~G.~M., \&
  {Bakes}, E.~L.~O. 1995, \apj, 443, 152

\bibitem[{{Wolfire} {et~al.}(2003){Wolfire}, {McKee}, {Hollenbach}, \&
  {Tielens}}]{WOLFIRE03}
{Wolfire}, M.~G., {McKee}, C.~F., {Hollenbach}, D., \& {Tielens}, A.~G.~G.~M.
  2003, \apj, 587, 278

\bibitem[{{Wong} \& {Blitz}(2002)}]{WONG02}
{Wong}, T., \& {Blitz}, L. 2002, \apj, 569, 157

\bibitem[{{Yoshizawa} \& {Noguchi}(2003)}]{YOSHIZAWA03}
{Yoshizawa}, A.~M., \& {Noguchi}, M. 2003, \mnras, 339, 1135

\bibitem[{{Young} {et~al.}(1996){Young}, {Allen}, {Kenney}, {Lesser}, \&
  {Rownd}}]{YOUNG96}
{Young}, J.~S., {Allen}, L., {Kenney}, J.~D.~P., {Lesser}, A., \& {Rownd}, B.
  1996, \aj, 112, 1903

\bibitem[{{Young} {et~al.}(1986){Young}, {Schloerb}, {Kenney}, \&
  {Lord}}]{YOUNG86}
{Young}, J.~S., {Schloerb}, F.~P., {Kenney}, J.~D., \& {Lord}, S.~D. 1986,
  \apj, 304, 443

\bibitem[{{Young} \& {Scoville}(1991)}]{YOUNG91}
{Young}, J.~S., \& {Scoville}, N.~Z. 1991, \araa, 29, 581

\end{thebibliography}


\end{document}